%% file: paper.tex
\documentclass{ametsocV6.1}
\nolinenumbers
\let\internallinenumbers\relax
\usepackage[percent]{overpic}  

\title{Generative and deterministic deep learning models comparison for fine-scale precipitation retrievals from infrared brightness temperature}

\authors{
Matthieu Meignin,\aff{a,b}\correspondingauthor{Matthieu Meignin, matthieu.meignin@latmos.ipsl.fr}  
Cécile Mallet,\aff{a,b} and 
Nicolas Viltard\aff{a}\thanks{This Work has been submitted to \textit{Artificial Intelligence for the Earth Systems}. Copyright in this Work may be transferred without further notice.}
}

\affiliation{
\aff{a}{LATMOS/IPSL, UVSQ Université Paris-Saclay, Sorbonne Université, CNRS, Guyancourt, France} \\
\aff{b}{ARCHES Team, Inria, Paris, France}
}

\abstract{Accurate precipitation estimation at fine spatial scales is critical for hydrology, agriculture, and climate studies. Infrared brightness temperatures from geostationary satellites offer excellent temporal coverage over continental-scale domains. However, because these measurements primarily characterize cloud-top properties rather than precipitation processes near the surface, their correlation with rainfall intensity remains limited, making quantitative precipitation estimation challenging. In this study, we conduct a systematic inter-comparison of state-of-the-art deep learning models for high-resolution precipitation retrieval from Meteosat Second Generation infrared brightness temperatures over metropolitan France. These models include deterministic U-Nets, transformer-based architectures, conditional GANs, and diffusion models. \\
We construct a curated dataset spanning 2008–2023, combining Météo-France radar mosaics as reference with multi-channel infrared observations, and design preprocessing and sampling strategies to address the heavy-tailed, intermittent nature of rainfall. Our results show that deterministic models provide robust mean estimates and excel in pixel-wise accuracy, but systematically underestimate extreme precipitation. In contrast, generative models better capture the full precipitation distribution, including rare and heavy rainfall events, producing more realistic spatial structures at the cost of reduced pixel-wise fidelity.  \\
These results highlight a trade-off between pixel-wise accuracy and precipitation variability, showing that generative approaches are advantageous for extreme-event detection and probabilistic applications. This work establishes a reproducible framework for evaluating infrared-based precipitation retrieval methods and provides guidance for designing models that balance precision, variability, and extreme-event representation. }

\begin{document}

\maketitle

%
%
%
\statement
This study provides a systematic comparison of deterministic and generative deep learning models for precipitation estimation from infrared satellite observations. By quantifying the trade-offs between pixel-wise accuracy and capturing extreme rainfall events, it offers a reproducible framework and methodological insights that guide the design of precipitation retrieval models, supporting operational rainfall estimation and extreme-event assessment.

%
\section{Introduction}

Precipitation estimation is essential for a wide range of applications, including agricultural planning, water resource management, disaster mitigation, and climate research. Traditionally, precipitation retrieval relies on ground-based observations, such as gauges and radar networks. However, such gauge networks are too sparse and lack the spatial coverage to capture the spatial variability of rain at global scale. On the other hand, radar networks are widely used for precipitation retrieval over land surfaces, although their coverage is very country-dependent. Thus, satellite-based observations are essential to overcome these coverage limitations.

Among the available satellite sensors, Low Earth Orbit (LEO) satellites with Passive Microwave Radiometers (PMR) are widely used in order to perform quantitative precipitation estimation via remote sensing. Microwave sensors provide an indirect but strongly correlated measurement of rainfall, yet LEO satellites offer limited spatial and temporal coverage. To surpass this inherent limitation of the microwave measurement, we aim to leverage satellite sensors with higher spatial and temporal coverage. Infrared imagers aboard geostationary platforms observe nearly one-third of the Earth’s surface at 10 to 30 minute intervals. Although infrared imagers provide valuable information on cloud properties, their correlation with surface precipitation is generally weaker than that of PMR.

Traditional methods to retrieve precipitation from infrared data go back to the 1980s, primarily as indicators of convective activity, especially in tropical regions~\citep{Arkin_Ardanuy1989}. In the past two decades, the ability of artificial neural networks to model complex non-linear functions has encouraged their use in precipitation estimation from infrared imagers. In~\citet{Yang_Hong2004}, a segmentation of infrared geostationary satellite imagery is performed with a self-organizing feature map. More recently, convolutional neural networks (CNNs) have been applied for infrared precipitation retrieval. In~\citet{C_Wang2021} and~\citet{Dongling_Wang2024}, convolutional networks were used to perform infrared precipitation retrieval over China.~\citet{Berthomier2023} validated the use of the DeepLabV3+ architecture for rainfall retrieval over metropolitan France and its overseas territories.
In~\citet{Hayatbini2019} and~\citet{Han2025} conditional generative adversarial networks (cGANs) were employed to improve precipitation retrieval by leveraging generative deep learning. In \citet{Amell2025}, a U-Net-inspired architecture was used to perform infrared precipitation retrieval over Africa. By exploiting extensive datasets of infrared brightness temperature (BT) observations, deep learning approaches can be trained to approximate the complex, non-linear precipitation retrieval function that maps infrared radiances to rain rates.

Despite these promising advances, direct comparison of results across studies remains challenging due to divergent approaches and study settings. First, the geographic diversity of the studied regions (e.g., France, China, Africa) introduces significant variability in meteorological characteristics and precipitation regimes, with each study sampling different weather conditions and seasonal patterns. Second, evaluation metrics are not standardized across publications, making quantitative performance comparisons difficult. Third, reference precipitation targets vary considerably, ranging from ground-based radar networks and spaceborne radar observations to satellite-derived precipitation products, each with distinct uncertainties and spatial-temporal resolutions. Fourth, the infrared imagers employed differ across studies, introducing sensor-specific biases and measurement characteristics. These inconsistencies prevent meaningful inter-comparison and limit our ability to draw robust conclusions about optimal methodological choices for infrared-based quantitative precipitation estimation (QPE). To our knowledge, no study has yet compared deterministic and generative deep learning architectures for infrared-based precipitation retrieval within a single, consistent framework.

This study aims to address this methodological gap by conducting a systematic inter-comparison of deep learning models and strategies that have shown promising results in the literature. We evaluate several state-of-the-art architectures, including U-Net variants (baseline U-Net, Multitask U-Net, and Quantile regression U-Net), convolutional and transformer-based models (DeepLabV3+ and SegFormer), conditional GANs, and diffusion models (DDPM), all trained and tested on the same dataset with consistent evaluation metrics and reference targets. By standardizing the experimental framework, we seek to identify the essential modeling components and architectural choices that are critical for accurate QPE from infrared observations. Although the present work focuses on deep learning model inter-comparison rather than operational product evaluation~\citep{Nguyen2020,Huffman2020}, the results are intended to inform the development of future satellite precipitation retrieval chains with fine-scale reconstruction capabilities.

 We also evaluate how different approaches deal with the specific challenges inherent to precipitation estimation. Challenges in precipitation retrieval go beyond the indirect relationship between cloud-top brightness temperatures and surface rainfall. They stem from the statistical properties of precipitation fields: high intermittency (frequent zero values), heavy-tailed distributions (extreme values deviate strongly from Gaussian behavior), multifractal scaling, and scale-dependent relationships. The rarity of extreme precipitation events, which are nonetheless critical for hydrological applications such as flood forecasting, poses specific difficulties during model training due to class imbalance. These characteristics of precipitation necessitate careful consideration of loss functions, evaluation metrics, and modeling strategies, as traditional approaches optimized for mean squared error may fail to adequately represent the full distribution of precipitation, particularly its tail behavior and spatial variability across scales. For this reason, model performance is analyzed separately across precipitation intensity ranges, from no rain to violent rainfall.

This study focuses on France for three main reasons. First, we have access to a large, high-quality database suitable for deep learning applications, spanning the period from 2008 to 2023, which enables robust model training and evaluation across a wide range of meteorological conditions. Second, France exhibits a large diversity of precipitation regimes, ranging from large-scale stratiform precipitation associated with winter frontal systems to highly localized convective events such as summer thunderstorms, providing a stringent test bed for infrared-based precipitation estimation methods. Finally, France benefits from a dense ground-based radar network, allowing the use of high-resolution radar products as reference data at 1 km. This enables us to train and evaluate deep learning models against a fine-scale reference, and to assess the extent to which coarser infrared observations can inform precipitation structures at that scale.
Using this long-term, high-resolution dataset, the present study provides a comprehensive and controlled evaluation of modern deep learning approaches for infrared-based quantitative precipitation estimation, with the objective of identifying modeling strategies that are robust across diverse rainfall regimes.

\newpage

\section{Study Domain and Data Sources}

\subsection{Study area}

\begin{figure}[h]
  \centering
 \noindent\includegraphics[width=19pc,angle=0]{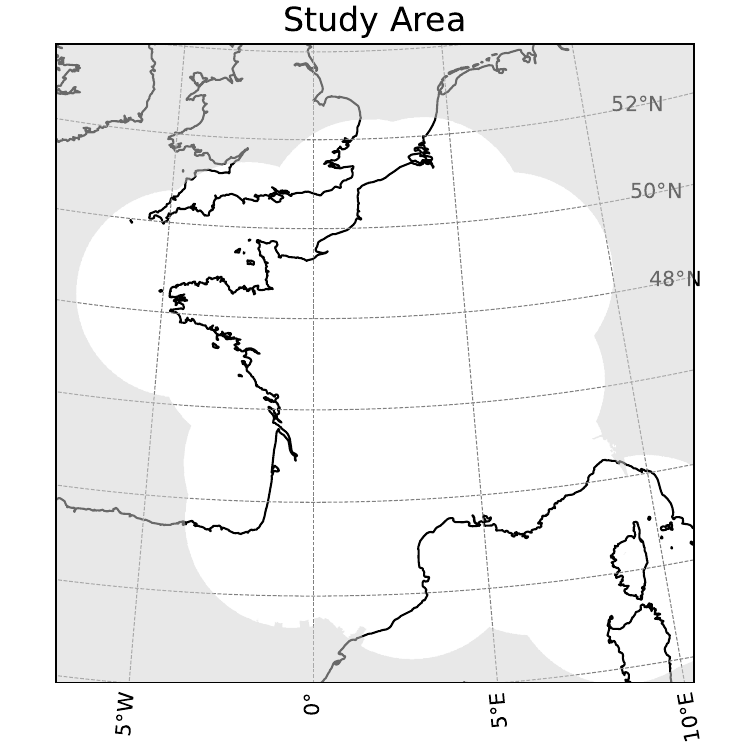}\\
  \caption{Study area covering longitudes from 10° W to 14° E and latitudes from 39° N to 54° N. Light
grey areas indicate regions without ground truth data for training.}\label{f1}
\end{figure}

We focused on the metropolitan France region from 10° W to 14° E and 39° to 54° N with a spatial resolution of 1 km. Although the study area is relatively small, the 1 km spatial resolution allows evaluation of the model’s performance at a fine scale and the 15-year radar archive provides a reliable ground truth for training.
The region experiences predominantly mid-latitude precipitation events, which are generally less intense than tropical convection. These events can be challenging for infrared-based retrievals due to weaker cloud–rain correlations.
Figure~\ref{f1} shows the geographical extent and the location of available radar observations. Light grey areas indicate regions outside of radar coverage.

\subsection{Input data: Meteosat Second Generation infrared Brightness Temperature (BT)}

To perform precipitation estimation from geostationary satellites, we use infrared brightness temperatures acquired by the Spinning Enhanced Visible and InfraRed Imager (SEVIRI) onboard Meteosat Second Generation (MSG). The instrument observes the Earth in 11 spectral channels with a 15-minute repeat cycle~\citep{Schmetz2002}. The geostationary satellite is located at approximately 0° longitude. This enables continuous monitoring of our domain at high temporal resolution. For this study, we use SEVIRI data spanning 2008 to 2023, providing a long-term record suitable for training, validation, and testing.

We select the following channels: 6.2 µm and 7.3 µm to detect water vapor at different levels of the troposphere, 8.7 µm for thin cirrus clouds and to support the discrimination between ice and water clouds, 9.7 µm for tracking ozone patterns representative of wind motion in the lower stratosphere, 10.8 µm and 12.0 µm for measuring sea and land surface as well as cloud-top temperatures, and 13.4 µm for cloud-top height. We do not use the 3.9 µm channel, even though it contains information for cloud and fog detection, because of solar contamination during daytime~\citep{Krebs2007}. The characteristics of the infrared channels used in this study are shown in Table~\ref{tab:seviri_channels}.

\input{msg_file_format}

The SEVIRI channels have an effective spatial resolution of 3 km at nadir, but due to the off-nadir viewing geometry over metropolitan France, this degrades to approximately 6 km~\citep{EUMETSAT2017}. Consequently, although the retrieval grid is defined at 1 km resolution, the infrared observations remain inherently coarser, with a spatial resolution of about 6 km over France. As a result, small-scale precipitation features cannot be directly resolved from the satellite signal alone. The 1 km formulation should therefore be interpreted as a precipitation inversion and statistical downscaling framework conditioned on infrared observations, consistent with the native resolution of the radar reference data used for training, rather than as a claim of true retrieval accuracy at 1 km. The ability of the models to reconstruct fine-scale structures ultimately depends on the information content available in the infrared observations.

\subsection{Target data: Météo-France Radar Mosaic}

The target data used in this study consist of 5-minute accumulated rainfall estimates derived from the Météo-France radar mosaic~\citep{ParentDuChatelet2003, meteofrance2019}. Each mosaic covers metropolitan France with 1536×1536 pixels at 1 km resolution. The 5-minute accumulation values are converted to mm/h by multiplying by a factor of 12. Rainfall values are quantized with a precision of 0.12~mm~h$^{-1}$.

Each pixel has a dynamic quality index from 0 (unusable) to 100 (excellent), representing measurement reliability. For each pixel, the index reflects the best quality among contributing radars.

These mosaics provide a reliable ground truth for training deep learning models and evaluating precipitation estimates at fine spatio-temporal scales. Any pixel with a quality index below the selected threshold is excluded from the training, validation, and test datasets.

\section{Training Data Construction}
\label{training_data_constr}

Preprocessing and training data construction are critical for precipitation estimation using deep learning. Instantaneous rainfall fields exhibit strong intermittency and heavy-tailed distributions, characterized by a predominance of zero or near-zero values and rare but intense events. If training samples are selected naively, learning algorithms tend to be dominated by no-rain conditions, leading to biased predictions and poor representation of moderate and heavy precipitation.

To mitigate these issues, we construct a curated database designed to expose models to the full range of rainfall intensities, spatial structures, and temporal variability observed over the study domain. This procedure also enables the removal of unreliable measurements, consistent normalization of inputs, and the application of augmentation strategies that improve training stability and generalization.

\subsection{Collocations and Tiling}

Radar-derived precipitation estimates from the Météo-France 5-minute mosaics are collocated with SEVIRI infrared brightness temperatures. SEVIRI full-disk images are acquired every 15 minutes, with metropolitan France observed during a fixed scan window in each cycle. Each infrared image is assigned an effective acquisition time corresponding to the center of this scan window over France, and is then matched to the nearest 5-minute radar accumulation. Spatial collocation is performed by resampling the SEVIRI data to the Météo-France radar grid using a nearest-neighbor approach.

The collocated fields are divided into non-overlapping tiles of 256×256 pixels. This tiling ensures that each infrared observation is paired with a spatially coherent reference precipitation field while preserving fine-scale rainfall structures.

\subsection{Event Selection and Data Split}
\label{sec:event_selection}
To ensure temporal independence between training samples, consecutive rainfall events are separated by a minimum of two hours. This avoids over-representing persistent systems and limits temporal autocorrelation between samples.

Because heavy rainfall events are rare but essential for model performance, we explicitly address class imbalance through targeted oversampling. We oversample tiles containing sufficient numbers of low, medium, or high rain pixels ($\ge$ 0.5, 1, 5 mm/h), retaining tiles that meet thresholds of 20,000, 1,000, or 200 pixels respectively. Thresholds were empirically determined; future studies could optimize these values further.

Training and validation samples are drawn from different days spanning 2008–2023, while the entire year 2019 is reserved for independent testing. This results in approximately 80,000 training tiles, 20,000 validation tiles, and 1,349 test tiles.

\subsection{Quality Control and Normalization}

We use the Météo-France radar quality index to filter unreliable reference data, retaining only pixels with a quality score above 65. This threshold was determined empirically through exploratory analysis, as lower values often correspond to beam blockage, clutter, or attenuation artifacts. Pixels below this threshold are treated as missing values. Infrared brightness temperatures are normalized by subtracting the training-set mean and dividing by three times the training-set standard deviation. This standardization is preferred over min–max scaling, as it allows the models to handle future observations that may fall outside the range encountered during training, which is particularly relevant for long-term climate datasets.

\subsection{Augmentation and Target Transformation}
To increase sample diversity and improve generalization, random horizontal and vertical flips are applied during training. We evaluated random crop sizes between $128\times128$ and $224\times224$ pixels and found the optimum to be architecture-dependent: the U-Net family performs best with $128\times128$ crops, with performance degrading at $224\times224$, whereas DeepLabV3+ and SegFormer perform best with $224\times224$ crops and degrade at $128\times128$. Each architecture is therefore trained at its preferred crop size; the per-model settings are reported in Appendix~B. These augmentations reduce overfitting.

Several transformations of the target precipitation fields were explored to account for their heavy-tailed distribution. Although logarithmic transformations are commonly used in precipitation retrieval~\citep{Amell2025, Han2025, rs16162871} and nowcasting studies~\citep{Geng2023}, we did not observe significant performance gains in our experiments; applying a logarithmic transformation to train U-Net with a mean squared error even degraded performance on heavy and violent precipitation. Instead, inspired by the zero-noising approach of~\citet{Amell2025}, we inject noise at zero-rainfall pixels to improve rain/no-rain discrimination, which is particularly important given the high frequency of zero-rain pixels.

\section{Methodology}

We evaluate a range of deep learning models for precipitation estimation, focusing on two main paradigms: deterministic and generative. Deterministic approaches estimate specific functionals of the conditional distribution $p(Y \mid X)$, such as the mean, median, or class probabilities, whereas generative models learn a probabilistic representation of either the joint distribution $p(X,Y)$ or the conditional distribution $p(Y \mid X)$, enabling the generation of realistic samples that reflect precipitation variability. This distinction fundamentally affects how precipitation estimates are represented and interpreted.

Deterministic models learn a direct mapping from satellite observations to pixel-wise precipitation estimates. They typically optimize a loss function such as Mean Squared Error (MSE) or Mean Absolute Error (MAE), producing the conditional mean or median, respectively, for each pixel. While straightforward to train and evaluate, deterministic models tend to underestimate extreme rainfall and smooth fine-scale spatial structures, limiting their ability to capture the highly variable and intermittent nature of precipitation.

Generative models~\citep{Ho2020DDPM, Isola2017Pix2Pix}, on the other hand, learn the full conditional distribution. By sampling from this distribution, they enable the generation of realistic samples that reflect the variability of the data. These models are generally more complex and computationally demanding, and evaluating their outputs is less straightforward, as plausibility must be assessed alongside statistical fidelity.

The skewed and zero-inflated distribution of rainfall makes generative modeling particularly relevant. Most pixels are dry or lightly raining, while extreme events form a heavy tail that is critical for hydrological and climate applications. Careful training and validation strategies are required to avoid overfitting, artifacts, or physically implausible “hallucinations.”

The following subsections describe each model in detail, while implementation details are deferred to Appendix~~B.

\subsection{Deterministic}

In this section, we present models trained using pixel-wise metrics to produce deterministic precipitation estimates. Our goal is to evaluate overall performance and identify the best-performing deterministic models for fine-scale precipitation retrieval over France.

\subsubsection{U-Net}

The U-Net~\citep{Ronneberger2015UNet} is a convolutional neural network architecture widely used for image-to-image tasks. It consists of a contracting path with 3×3 convolutions, max-pooling, and ReLU activations, which progressively down-sample the input image into feature maps. The network then reconstructs the output through an expansive path using transposed convolutions for up-sampling. Skip connections between corresponding layers of the contracting and expansive paths allow the network to combine local and global information effectively.

\subsubsection{U-Net Variations}

In addition to the standard U-Net, which serves as the baseline model in this study, we explore several variations by modifying the loss functions. These variations are designed to investigate whether alternative training objectives can improve precipitation retrieval performance. The following subsections describe the specific U-Net variants considered in this work.

\paragraph{Quantile Regression U-Net}

The Quantile Regression U-Net uses the same architecture as the standard U-Net described above. The key difference lies in the training objective: rather than predicting a single value per pixel, the network is trained to predict multiple quantiles of the rainfall distribution. Estimating quantiles allows the model to capture the full conditional distribution of precipitation, which is particularly important given the highly skewed and intermittent nature of rainfall.

The training objective is defined following~\citet{Viltard2023, Amell2025} as the sum of quantile losses for each quantile $\tau$:

\begin{align}
L_\tau(y_i, \hat{y}_i) &=
\begin{cases}
\tau \, (y_i - \hat{y}_i), & \text{if } y_i - \hat{y}_i \ge 0 \\
(\tau - 1) \, (y_i - \hat{y}_i), & \text{if } y_i - \hat{y}_i < 0
\end{cases} \\[2mm]
L_\tau(y, \hat{y}) &= \frac{1}{N} \sum_{i=1}^{N} L_\tau(y_i, \hat{y}_i) \\[1mm]
\mathcal{L}(y, \hat{y}) &= \sum_{\tau} L_\tau(y, \hat{y})
\end{align}

Here, $\hat{y}$ represents the predicted value, $y$ is the observed target, $\tau$ denotes the quantile, and $N$ is the total number of pixels. In this study, the network is trained to predict the deciles of the rainfall distribution ($\tau = 0.1, 0.2, 0.3, 0.4, 0.5, 0.6, 0.7, 0.8, 0.9$). For evaluation, we extract the median prediction ($\tau$= 0.5) from the quantile outputs and use it as a single deterministic estimate. This provides a consistent point-wise comparison with baseline models that produce a single rainfall prediction.

\paragraph{Multitask U-Net}

The Multitask U-Net retains the standard U-Net architecture, with the key modification that the final layer is split into two heads: one for regression and one for segmentation. This design allows the network to simultaneously predict continuous rainfall values and binary precipitation occurrence, leveraging both tasks to improve overall performance. The combined loss function is defined as follows:

\textbf{1. Regression Loss (MSE):}
\begin{align}
\mathcal{L}_{\text{reg}} = \frac{1}{N} \sum_{i=1}^{N} (\hat{y}_{\text{reg},i} - y_{\text{reg},i})^2
\end{align}

\textbf{2. Segmentation Loss (BCE):}
\begin{align}
\mathcal{L}_{\text{seg}} = - \frac{1}{N} \sum_{i=1}^{N} \Big[ 
y_{\text{seg},i} \log \sigma(\hat{y}_{\text{seg},i}) + 
(1 - y_{\text{seg},i}) \log (1 - \sigma(\hat{y}_{\text{seg},i})) 
\Big]
\end{align}

\textbf{3. Combined Two-Head Loss:}
\begin{align}
\mathcal{L}_{\text{total}} = w_{\text{reg}} \, \mathcal{L}_{\text{reg}} + w_{\text{seg}} \, \mathcal{L}_{\text{seg}}
\end{align}

Here, $\hat{y}_{\text{reg},i}$ and $y_{\text{reg},i}$ denote the regression prediction and target for pixel $i$, while $\hat{y}_{\text{seg},i}$ and $y_{\text{seg},i}$ represent the segmentation logit and binary target (1 if precipitation $>0.2$ mm/h). $\sigma(x)$ is the sigmoid function, and $w_{\text{reg}}$ and $w_{\text{seg}}$ are the weights for the regression and segmentation losses, respectively (set to 1 and 10), chosen so that the two loss terms are of comparable magnitude during training.
This multitask design allows the network to leverage both regression and classification signals.

\subsubsection{DeepLabV3+}

DeepLabV3+~\citep{Chen2018EncoderDecoder} is a widely adopted architecture for image segmentation tasks and is well-suited to our precipitation retrieval problem. We utilize a lightweight version of its Atrous Spatial Pyramid Pooling (ASPP) module, which reduces the risk of overfitting compared to larger modules. 

The ASPP module aggregates features at multiple spatial scales, which is particularly relevant for precipitation fields exhibiting multiscale and multifractal properties~\citep{Verrier2011}. By capturing structures from small convective cells to large-scale frontal systems, the model can better learn the variability of rainfall patterns. 

Atrous (dilated) convolutions within the module enlarge the receptive field without increasing the number of parameters, allowing the network to integrate global context and spatially extended rainfall structures. This is important for representing coherent precipitation systems that span tens to hundreds of kilometers.

\subsubsection{SegFormer}
SegFormer~\citep{Xie2021SegFormer} is a semantic segmentation model that combines a hierarchical Transformer encoder with an all-MLP decoder. The hierarchical encoder extracts features at multiple resolutions without requiring positional encoding, while the self-attention mechanism captures long-range spatial dependencies. This is particularly relevant for precipitation retrieval, where coherent systems such as frontal bands extend across tens to hundreds of kilometers. In this study, we use the MiT-B2 encoder, which offers a balance between model capacity and computational cost. Further details are provided in Appendix~~B.

\subsection{Generative Models}

Generative models go beyond predicting the conditional mean by learning the full distribution of precipitation, enabling the generation of realistic samples that reflect both spatial variability and extreme events. Two main approaches are considered: GANs, which learn the distribution implicitly via adversarial objectives, and diffusion models, which explicitly model a stochastic generative process through progressive noise injection and removal. Both approaches are particularly effective at capturing heavy rainfall and preserving realistic precipitation textures.

In this study, generative models are trained only on pixels with valid precipitation values, resulting in a slightly smaller training set compared to deterministic models. Tiles with less than 5\% missing pixels have those pixels interpolated, while tiles exceeding this threshold are discarded. After preprocessing, the dataset comprises approximately 50,000 training tiles and 10,000 validation tiles, with the testing set unchanged. Although this reduces the amount of available data relative to the deterministic models, it is a necessary constraint: unlike deterministic models, which can handle missing pixels through selective backpropagation without learning a complete data distribution, generative models require fully valid inputs to avoid corrupting the learned distribution and destabilising training. This asymmetry therefore reflects an inherent limitation of the generative paradigm under real-world data quality conditions, rather than an arbitrary experimental choice, and should be considered when interpreting comparisons across model families. To ensure a fair comparison, all deterministic models were retrained on the restricted dataset used for the generative models, confirming that the performance gap remains consistent with the results presented here (Appendix~~E).

\subsubsection{Conditional GANs}

Conditional Generative Adversarial Networks (cGANs) extend the standard GAN framework by conditioning both the generator and discriminator on additional information, such as satellite observations, allowing the model to learn a mapping from input features (infrared brightness temperature) to precipitation fields. Following ~\cite{Hayatbini2019} and~\cite{Han2025}, we employ the Pix2Pix architecture of ~\cite{Isola2017Pix2Pix}, consisting of a U-Net generator and a PatchGAN discriminator. This configuration serves as our baseline conditional GAN for precipitation retrieval.

GANs can suffer from mode collapse, so it is important to balance the adversarial loss with a reconstruction loss such as L1 (mean absolute error) or L2 (mean squared error). After testing different configurations, we found that a combination of L1 loss and GAN loss provided the best performance for our dataset. Although cGANs belong to the generative model family by design, in our experiments we do not sample multiple latent codes: the generator receives a fixed conditioning input without stochastic perturbation, so the model produces a single output per input. However, the adversarial training objective fundamentally distinguishes the cGAN from deterministic regression: rather than minimizing a pointwise loss toward the conditional mean, the generator is trained to produce outputs indistinguishable from real precipitation fields under the discriminator. This encourages predictions that lie on the manifold of realistic precipitation fields rather than at the blurred conditional mean. Even when drawing a single output, the prediction preserves spatial texture and extremes rather than regressing toward a smoothed average.

\subsubsection{Diffusion Model}

Diffusion models are a class of generative models that learn to generate data by iteratively reversing a noise corruption process. In this study, we use a diffusion model based on the Denoising Diffusion Probabilistic Model (DDPM) framework~\citep{Ho2020DDPM}, employing an Attention U-Net as the denoising network and a cosine scheduler for the noise variance. 
During training, precipitation fields are gradually corrupted with Gaussian noise, and the network is trained to reconstruct the original data.
We experiment with two variants that share the same forward diffusion process but differ in their prediction target. In the \textit{epsilon-prediction} variant, the network is trained to predict the noise $\epsilon$ added at each step, from which the clean field is recovered by subtraction. In the \textit{image-prediction} variant, the network is instead trained to directly predict the clean precipitation field at each step. Both approaches are mathematically equivalent at the optimum but differ in practice in terms of training stability and sample quality.

For conditional precipitation estimation, the infrared brightness temperature channels are concatenated with the noisy input at each denoising step. This allows the model to generate precipitation fields conditioned on the satellite observations, producing high-quality samples that capture both the variability and the complex spatial structure of rainfall events.

A single generated sample is a draw from the learned conditional distribution $p(y \mid x)$: it preserves realistic spatial texture and extremes. Multiple samples can be drawn for the same conditioning input to form an ensemble~\citep{Guilloteau2025}. Averaging the ensemble approximates the conditional mean $\mathbb{E}[Y \mid X]$; this estimator behaves like a deterministic L2 predictor, improving pixel-wise scores while smoothing out the fine-scale structure and extremes that individual samples reproduce. Throughout Section~\ref{results} we therefore report realisation-wise metrics, computed from a single sample per input, while metrics obtained from the ensemble mean are reported in Appendix~D.


\section{Results}
\label{results} 

In this section, we present a systematic inter-comparison of the precipitation estimation models introduced earlier. We evaluate their performance across multiple perspectives:

\begin{enumerate}
    \item \textbf{Quantitative metrics} – We first assess model performance at the pixel level using a combination of binary rain/no-rain detection metrics: namely Accuracy, Precision, Recall, F1-score, Critical Success Index (CSI), False Alarm Ratio (FAR), and continuous precipitation intensity metrics, including the root mean square error (RMSE), mean absolute error (MAE), and the Pearson correlation coefficient. These metrics provide a rigorous and widely adopted baseline for inter-model comparison. However, as they primarily focus on pointwise errors, they naturally tend to favor deterministic models trained to minimize pixel-wise loss functions, and may not fully reflect a model’s ability to represent uncertainty or complex distributional properties.
    \item \textbf{Qualitative analysis} – To complement the quantitative evaluation, we conduct a qualitative analysis based on two representative case studies designed to assess how well the models reproduce the spatial organization of precipitation. The first case corresponds to a convective event, characterized by localized and intense rainfall structures, while the second case represents a stratiform precipitation event with broader and more homogeneous spatial patterns. This visual assessment allows us to examine the realism and coherence of the simulated precipitation fields, particularly in terms of spatial morphology and organization. Such an analysis is especially relevant given the expertise of the authors in precipitation systems and their spatiotemporal properties.
    \item \textbf{Distributional and spectral assessment} - Finally, we evaluate the ability of the models to reproduce the statistical distribution of rainfall intensities by comparing the empirical distributions of retrieved and observed precipitation through corresponding histograms. This analysis focuses on the representation of the full range of precipitation values, including extremes and low-intensity events. Unlike pixel-wise error metrics, this distributional evaluation is particularly informative for generative models, which are explicitly trained to approximate the underlying data distribution rather than to optimize pointwise accuracy. We further complement this analysis with the omnidirectional spatial power spectral density  \citep[PSD;][]{ravuri2021skilful,Guilloteau2025}, which describes how precipitation variability is distributed across spatial scales, independently of the orientation of rain structures (i.e., averaged over all directions). An overly smooth field, for instance, lacks variability at small scales. The PSD thus provides a direct measure of structural realism.
\end{enumerate}

All models are evaluated on a separate test set consisting of unseen data, the entire year of 2019 that was not used during training or validation. The test set was extracted following the same procedure as the training and validation sets and is therefore biased toward higher precipitation events. This represents 1,349 tiles of 256x256 $\mathrm{km}^2$ throughout France.

\subsection{Precipitation Detection Performance}

We first assess model performance at the pixel level, using both binary rain/no-rain detection and continuous precipitation intensity metrics.
Table~\ref{tab:classification_metrics} summarizes rain/no-rain detection performance at a 0.2 mm/h threshold. Deterministic models demonstrate strong classification performance, with the Baseline U-Net already achieving high accuracy. More complex architectures (DeepLabV3+, SegFormer) yield minimal or no improvement. Among deterministic models, the Quantile Regression U-Net attains the highest accuracy and precision, while the Multitask U-Net excels in F1 score and CSI, highlighting the trade-off between correct detection and false-alarm rates. SegFormer achieves the highest recall but at the cost of overprediction.
\input{classification}

These results reflect the deterministic training objective (L2 loss), which directly minimizes pixel-wise errors and naturally favors higher classification metrics. In contrast, generative models (cGAN, diffusion) exhibit lower classification scores, as expected: their training objectives prioritize sampling from conditional distributions and generating realistic spatial structures rather than achieving exact pixel-wise matches.

\subsection{Pixel-wise Regression Evaluation} 

Deterministic models consistently outperform generative approaches in RMSE across all rainfall intensity classes, with the Multitask U-Net showing the lowest overall errors (Table~\ref{tab:rmse_metrics}).

The Quantile Regression U-Net excels at predicting zero-rain pixels but slightly underestimates moderate-to-heavy rainfall due to its MAE-like behavior. Generative models show higher RMSE, reflecting the conditional variability carried by individual realisations; averaging multiple samples brings the diffusion model closer to the deterministic scores, though it does not match them exactly (Appendix~D). Critically, all models exhibit saturation for violent rainfall (\(>50\) mm/h), revealing fundamental limitations of infrared brightness temperature as a predictor for intense precipitation events.
These results should be interpreted in light of the test set composition described in Section~\ref{training_data_constr}. By construction, the test set oversamples higher-precipitation events relative to their natural climatological frequency, which may slightly inflate the aggregate RMSE values reported in the ``Rain~$\geq$~0.2'' and ``All'' columns. However, as the category breakdown in Table~\ref{tab:rmse_metrics} shows, heavy and violent rainfall events still represent only 0.36\% and 0.014\% of the test set respectively, meaning that even after oversampling, extremes remain substantially under-represented relative to low-to-moderate intensity events. The effect on aggregate metrics is therefore modest. Furthermore, the oversampling strategy is identical across all models, ensuring that they are trained and evaluated under the same conditions.  The corresponding MAE, Pearson, and bias scores are reported in Appendix~~A. There, the Quantile Regression U-Net achieves the best MAE, since its quantile loss reduces to the MAE at the median. Pixel-wise evaluation, while naturally favoring deterministic models, remains a useful benchmark for generative models. It allows us to quantify how closely the predicted precipitation fields match observations at the finest spatial scale, providing a consistent baseline for comparison across all models. Although generative models are designed to capture the full conditional distribution and realistic spatial variability, deviations at the pixel level are expected and do not necessarily indicate poor performance in capturing the overall precipitation structure. Therefore, pixel-wise metrics such as RMSE, MAE, and correlation should be interpreted as one aspect of performance rather than a complete assessment of generative models’ capabilities.

\input{rmse}

\newpage

\subsection{Qualitative Analysis}

\begin{figure}[!tbp]
    \centering
    \includegraphics[width=\textwidth]{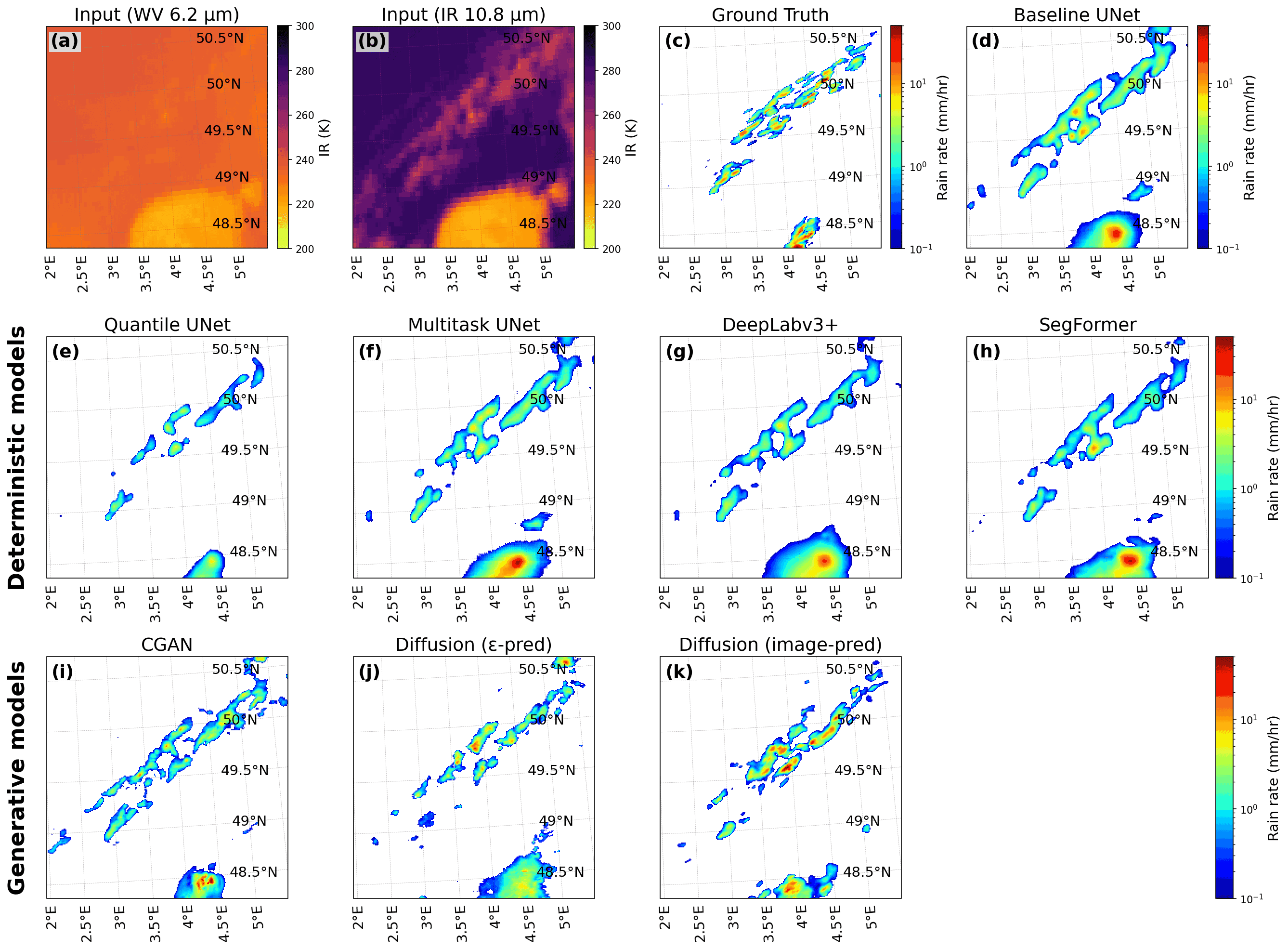}
    \caption{Case study of a convective precipitation event over the Grand Est region in France. The first row shows two of the model inputs, 6.2 µm and 10.8 µm infrared brightness temperatures, the ground-truth precipitation field, and the Baseline U-Net output. The second row presents precipitation estimates from all deterministic models, while the third row shows outputs from generative models (see Appendix Figures~C1 and~C2 for 16 additional samples from each diffusion variant). All panels correspond to the same time step and spatial domain.}
    \label{fig:convective}
\end{figure}

Case studies show that model performance is fundamentally limited by the information content of infrared observations. Figure~\ref{fig:convective} illustrates a convective event over the Grand Est region. Probable strong vertical motion produces distinctive cold cloud-top signatures, enabling the models to identify precipitating regions and generate plausible patterns for relatively small convective cells.

\begin{figure}[!tbp]
    \centering
    \includegraphics[width=\textwidth]{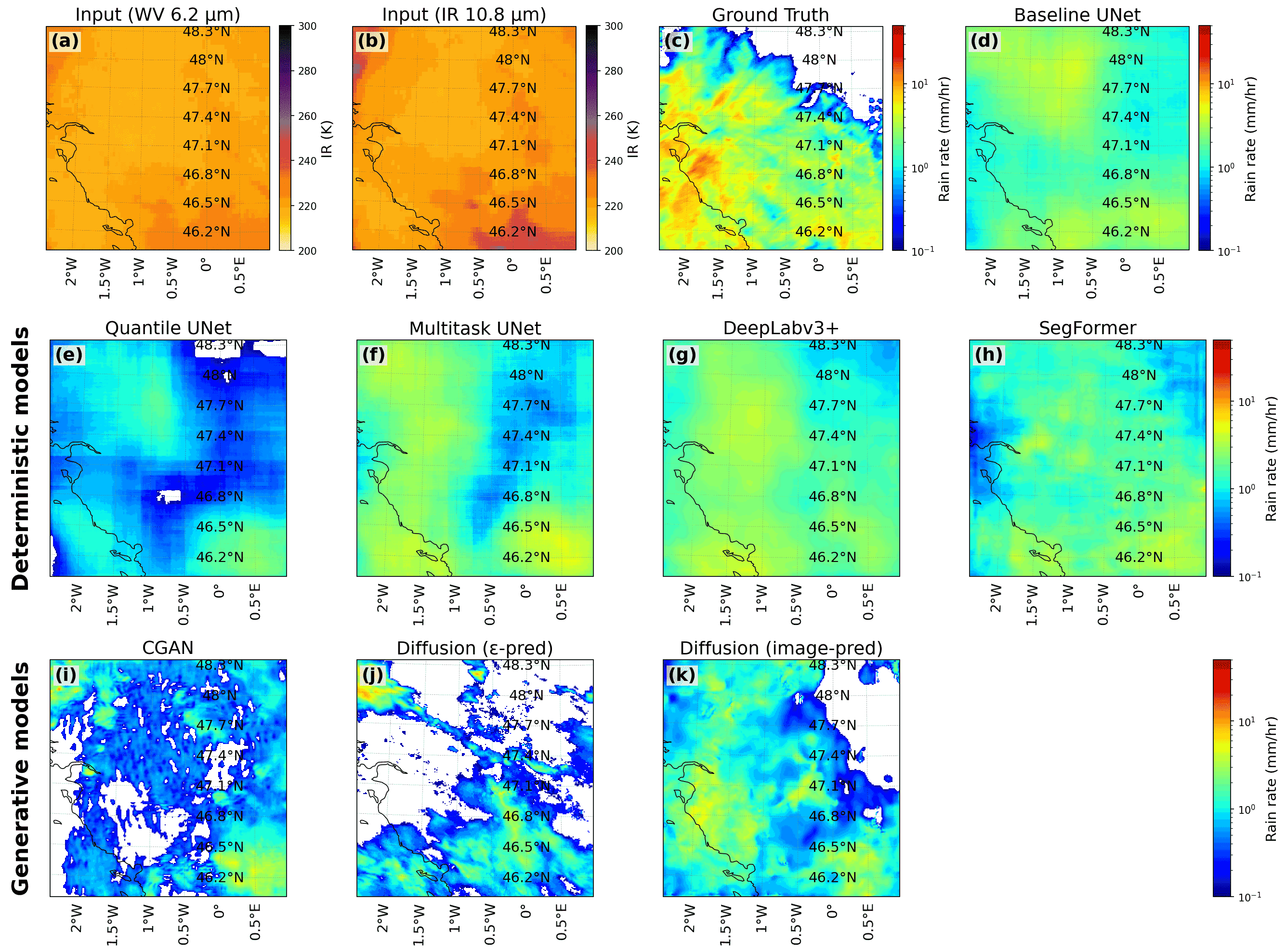}
    \caption{Case study of a stratiform precipitation event over the Pays de la Loire region in France. The first row shows two of the model inputs, 6.2 µm and 10.8 µm infrared brightness temperatures, the ground-truth precipitation field, and the Baseline U-Net output. The second row presents precipitation estimates from all deterministic models, while the third row shows outputs from generative models (see Appendix Figures~C4 and~C5 for 16 additional samples from each diffusion variant). All panels correspond to the same time step and spatial domain.}
    \label{fig:stratiform}
\end{figure}

Stratiform precipitation and larger convective systems are more challenging (cf. Figure~\ref{fig:stratiform}). Stratiform regions, composed of precipitation from decaying convective cells or situations with strong horizontal motion that laterally transport hydrometeors, exhibit cold infrared brightness temperatures similar to non-precipitating cirriform anvils. These anvil clouds persist after precipitation stops and contain ice particles that eventually evaporate, making them indistinguishable from precipitating stratiform regions in infrared observations alone. While microwave radiometers provide physically relevant information for distinguishing raining from non-raining regions and for characterizing precipitation intensity \citep{Viltard2023}, infrared observations, by contrast, are primarily sensitive to cloud-top radiative properties and cannot unambiguously discriminate between precipitating cores and non-precipitating anvil regions.

Stratiform events expose a shared limitation across all models: deterministic models produce overly smooth fields by regressing toward the conditional mean, while generative models produce spatially structured realisations that are not necessarily co-located with the observed precipitation. This spatial inconsistency is further illustrated in Appendix~C, where the ensemble of diffusion model realisations reveals high variability across realisations with limited spatial anchoring to the ground truth, in contrast to the convective case where realisations remain broadly consistent with the observed field. These results reflect both the inherent uncertainty in infrared-to-precipitation mapping and the fundamental trade-off between pixel-wise fidelity and distributional realism, which is further discussed in Section~\ref{results}~\ref{sectionDistrib}.

\newpage

\subsection{Distributional and Spectral Assessment}
\label{sectionDistrib}

\begin{figure}[!h]
    \centering
    \includegraphics[width=\textwidth]{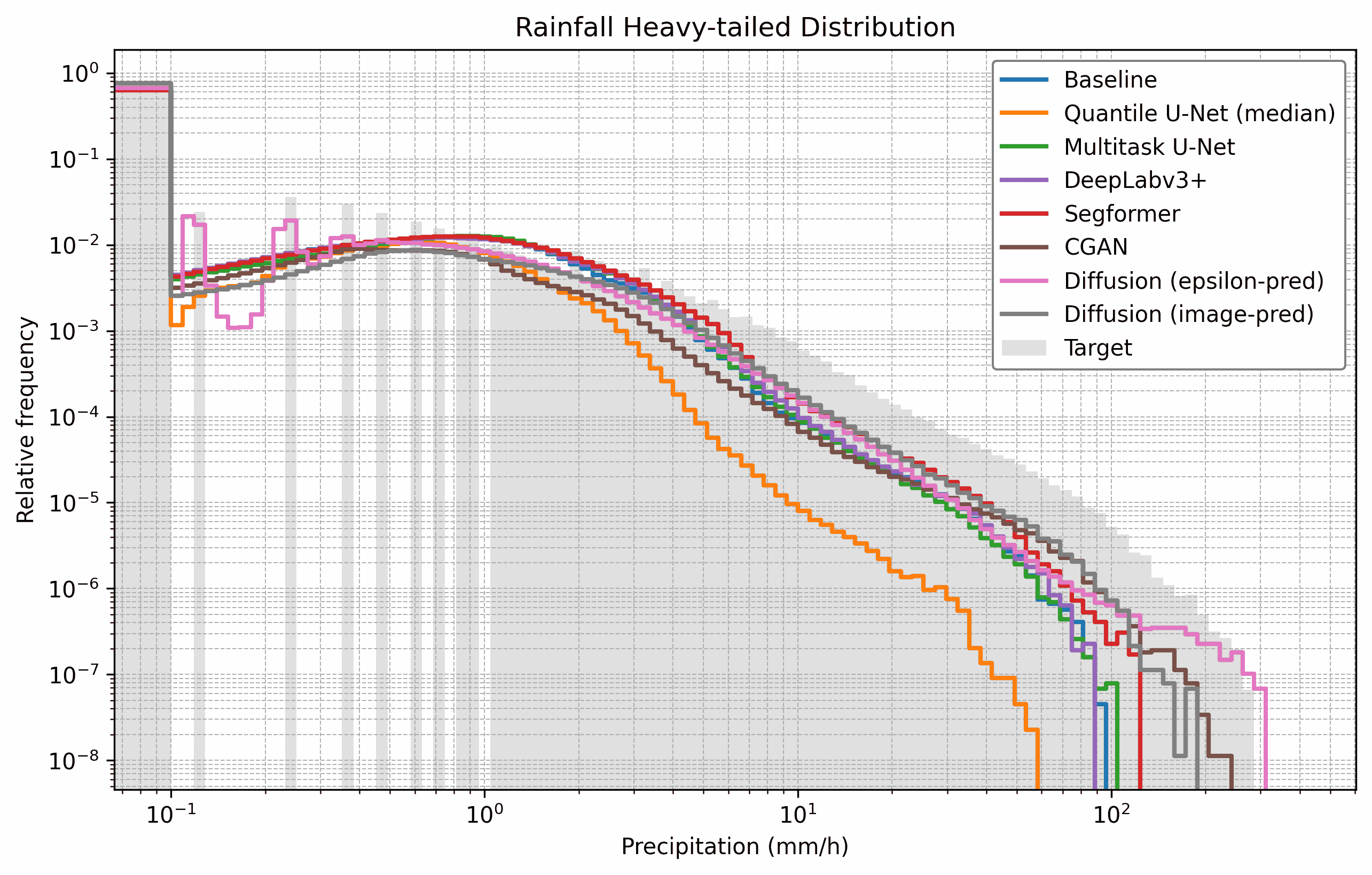}
    \caption{Histograms of observed and retrieved precipitation for the entire test set. Deterministic models tend to underestimate the heavy-rain tail, whereas generative models more accurately capture the full range of precipitation intensities.}
    \label{fig:histograms}
\end{figure}

To complement pixel-wise metrics, we assess how well the models capture the overall precipitation distribution. Figure~\ref{fig:histograms} shows the distributional performance of the models on the test set. Deterministic models reproduce the general precipitation distribution shape but systematically underestimate the heavy-rain tail, rarely producing values above 50 mm/h. While intense convective rainfall is often associated with deep cloud tops that are detectable in infrared observations, the relationship between cloud-top properties and surface rainfall remains indirect and uncertain. In addition, extreme precipitation exhibits much stronger spatial and temporal variability than stratiform rainfall, making the precise localization and intensity of convective cores difficult to infer from infrared observations alone. Deterministic training objectives further favor average solutions under such uncertainty, leading to a systematic underestimation of the heavy-rain tail.

To further assess the realism of spatial precipitation structures, we compute the omnidirectional spatial PSD over the test set following \citet{Guilloteau2025} and \citet{ravuri2021skilful}. Unlike pixel-wise metrics and distributional histograms, which evaluate pointwise intensity statistics, the omnidirectional PSD characterises how precipitation variability is distributed across spatial scales, independently of the orientation of rain structures.

\begin{figure}[!h]
    \centering
    \includegraphics[width=0.8\textwidth]{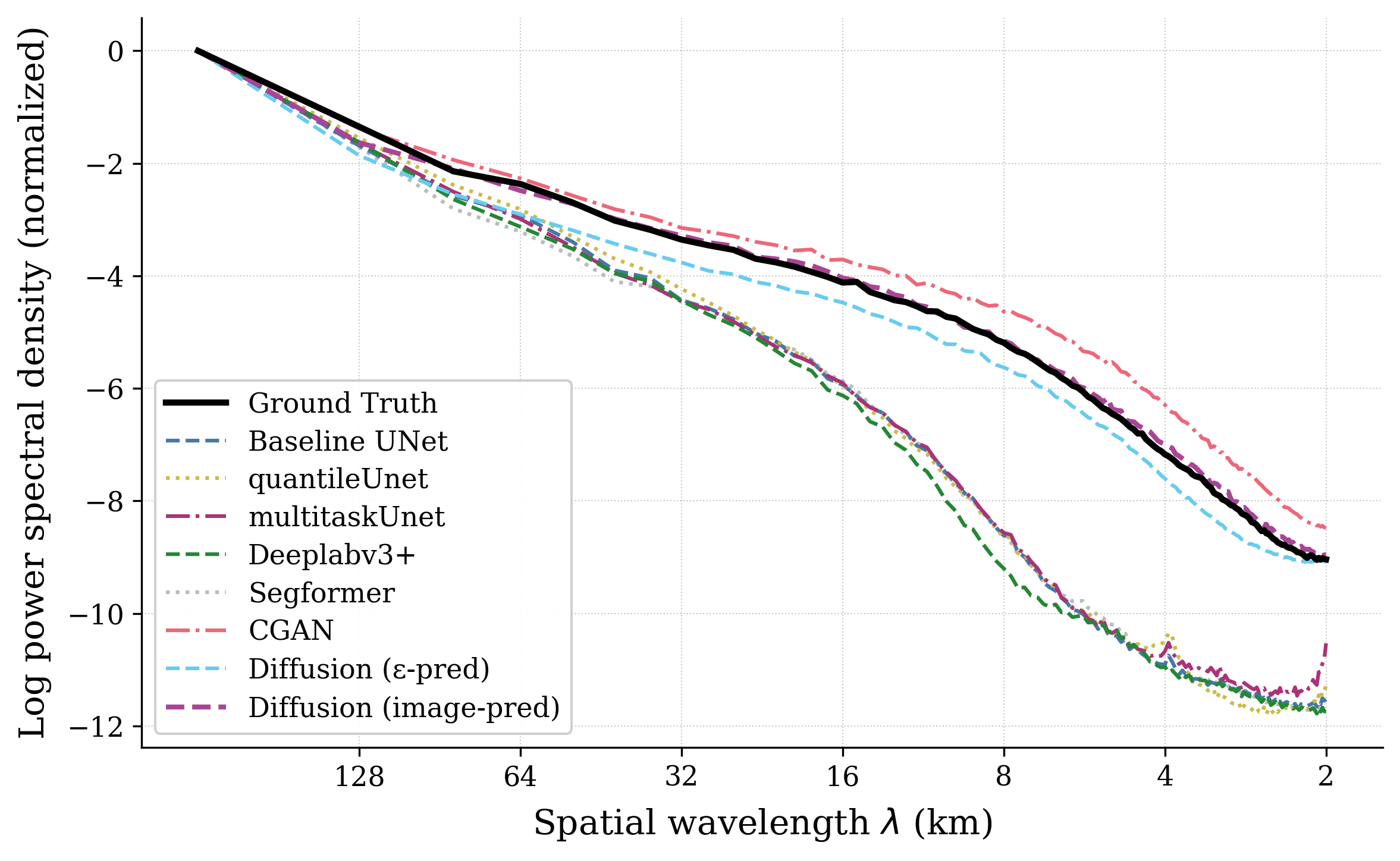}
    \caption{Omnidirectional spatial power spectral density (PSD) computed over the test set for each model alongside the ground truth (1~km spatial resolution).}
    \label{fig:psd}
\end{figure}

As shown in Figure~\ref{fig:psd}, deterministic models exhibit a systematic deficit of variability at small scales, consistent with the well-known smoothing effect of pixel-wise loss functions, which penalise deviations from the conditional mean and therefore suppress fine-scale variability. In contrast, generative models retain variability across all scales, with spectral slopes significantly closer to that of the ground truth, indicating that they produce precipitation fields with more realistic spatial structure.

Taken together, the distributional and spectral analyses confirm that the primary advantage of generative models (which model a conditional distribution rather than the conditional expectation) lies not in pixel-wise fidelity but in their ability to synthesize physically plausible, spatially structured precipitation fields consistent with observed rainfall statistics.

\newpage

\section{Discussion}

Our results highlight key trade-offs and operational considerations when applying deterministic and generative models for fine-scale precipitation retrieval from infrared brightness temperatures. Deterministic models reliably capture the conditional mean and produce consistent pixel-wise estimates, but they systematically underestimate extreme rainfall events and fail to fully represent spatial variability. In contrast, generative models, including cGANs and diffusion models, better reproduce the variability and heavy-rain tail of precipitation distributions, although they introduce additional challenges in training, evaluation, and inference.

Diffusion models, in particular the epsilon-prediction variant, tend to capture discretization patterns present in radar data, especially at low rain rates, as shown by the pink curve at low precipitation intensities in Figure~\ref{fig:histograms}. These artifacts of measurement precision provide limited physical information and can hinder the model’s ability to learn meaningful patterns.

Computational cost is a major consideration. Diffusion models require multiple iterative denoising steps, making inference substantially slower than all other methods presented in this study, which produce predictions in a single forward pass. Strategies such as reducing the number of denoising steps or using lightweight architectures can speed up inference but often compromise output fidelity or variability, limiting operational applicability. Emerging approaches using ODE solvers, such as flow matching~\citep{Lipman2023FlowMatching}, learn a continuous vector field that transforms noise into data. This enables sample generation with fewer iterations while maintaining performance comparable to standard diffusion models. Such techniques could provide a practical pathway to combine the advantages of generative modeling with operational efficiency.

From a practical standpoint, we observe that certain preprocessing and training choices have significant effects on model performance. Zero-noising of precipitation targets consistently improves rain/no-rain segmentation performance. By contrast, conventional normalization methods such as logarithmic scaling combined with min-max or mean-std normalization do not produce any noticeable improvement. Regarding loss functions, a hybrid approach combining binary cross-entropy (BCE) for classification and mean squared error (MSE) for regression achieves best overall performance. Quantile regression losses target specific percentiles of the conditional distribution: taking the median (50th percentile) tends to minimise MAE, while higher quantiles shift the focus toward extremes at the cost of increased MAE and reduced RMSE for typical rainfall conditions.

Overall, generative models provide the advantage of representing precipitation uncertainty and extreme events more realistically. However, practical deployment requires careful attention to coherence, quantization effects, and computational efficiency. Future work could explore hybrid architectures that combine deterministic precision with generative variability, or leverage multi-sensor data to improve the retrieval of extreme rainfall while maintaining operational feasibility.

This study establishes standardized methods for model comparison and evaluation, but several aspects would merit further investigation. We have focused on native-resolution performance (1 km × 1 km) without systematically assessing model behavior under spatial aggregation (e.g., 10 km or 25 km), which is relevant for hydrological and climate applications. Scale-dependent evaluation would clarify whether models preserve the correct scaling behavior of precipitation and whether relative performance rankings change with aggregation.

By addressing the statistical challenges inherent to precipitation such as intermittency, heavy tails, multifractality, and rare extreme events, this study provides a reproducible framework for future research in satellite-based precipitation estimation. Extending this framework to include scale-dependent metrics and geostatistical analyses would further strengthen model evaluation and support the development of operationally robust retrieval systems.

\newpage

\section{Conclusion}

This study provides a systematic intercomparison of state-of-the-art deep learning approaches for infrared-based precipitation retrieval under controlled and consistent conditions. We evaluated on the same dataset with standardized protocols a diverse set of models combining different architectures (convolutional and transformer-based) with both deterministic and generative training paradigms (including conditional GANs and diffusion). This analysis clarifies the influence of architectural choices, training strategies, and loss functions on retrieval performance.

The results reveal a fundamental trade-off between deterministic and generative approaches. Deterministic models excel in pixel-wise accuracy and provide reliable mean estimates, making them suitable for applications requiring precise spatial localization and low false-alarm rates. Generative models, by contrast, better capture precipitation variability and extreme events, making them more appropriate for probabilistic applications, ensemble forecasting, and scenarios requiring representation of the full range of rainfall patterns.

Both approaches are constrained by the inherent physical limits of infrared observations: the weak correlation between cloud-top brightness temperatures and surface precipitation, particularly for intense rainfall and stratiform systems. These limitations highlight the potential value of hybrid architectures that combine deterministic precision with generative variability, as well as the integration of complementary observations (e.g., microwave sensors) to improve extreme rainfall retrieval.

Overall, this study provides both practical guidance for operational infrared-based precipitation estimation and a framework for future research aimed at improving the representation of precipitation variability and extremes.

\newpage

\acknowledgments

This work was supported by the \textit{Terre-Océan-Surfaces Continentales-Atmosphère} program of the Centre National d'Etudes Spatiales (CNES) and the \textit{Programme National de Télédétection}. Computing and storage resources were provided by GENCI at IDRIS through Grant 2024-AD011012790R4 on the Jean Zay supercomputer's V100 partition.  

Additional support was provided by the France 2030 project \textit{Académie Spatiale d'Île-de-France} (\url{https://academiespatiale.fr/}), managed by the National Research Agency under reference ANR-23-CMAS-0041.

%
%
\datastatement
The SEVIRI Level 1.5 infrared brightness temperature data from Meteosat Second Generation (MSG) used in this study are freely available from EUMETSAT and can be accessed via the EUMETSAT Data Centre at \url{https://data.eumetsat.int/product/EO:EUM:DAT:MSG:HRSEVIRI}.  
The Météo-France 5-minute precipitation radar mosaics used as reference data are available from Aeris at \url{https://doi.org/10.25326/357}.


%



\appendix[A] 
\appendixtitle{Pixel-wise Performance Metrics}
\label{appendixA}

\input{bias}
\input{mae}
\input{pearson}

\newpage
\clearpage

\appendix[B]
\appendixtitle{Implementation details}
\label{appendixB}

This appendix details the architecture and training configuration of each model. All models were trained for up to 600 epochs, and the checkpoint at the best validation epoch (reported per model below) was retained for evaluation.

The optimal crop size was found to be architecture-dependent. The U-Net family was trained with $128\times128$ crops; increasing the crop size to $224\times224$ degraded its performance. Conversely, DeepLabV3+ and SegFormer were trained with $224\times224$ crops, as reducing the crop size to $128\times128$ degraded theirs. Several further settings were explored but not adopted. The preferential loss of \citet{Berthomier2023} was tested for the U-Net and DeepLabV3+ models: it improved performance on heavy and violent rain at the cost of light and moderate rain, and was not retained in order to preserve balanced performance across intensities. Regarding the logarithmic target transformation, removing it from the Quantile U-Net gave similar results, whereas applying it to the MSE-trained U-Net degraded performance on heavy and violent precipitation.

\input{implementation_unets}
\input{implementation_segformer_deeplav3+}
\input{implementation_generative}
\clearpage

\appendix[C]
\appendixtitle{Diffusion Model Ensemble Realisations}
\label{appendixC}
Unless otherwise specified, all figures in this appendix use the same colormap as the corresponding case study figures in the main article 
(Figures~\ref{fig:convective} and~\ref{fig:stratiform}).

\subsection{Convective Event}

\begin{figure}[!h]
\centering
\includegraphics[width=0.6\textwidth]{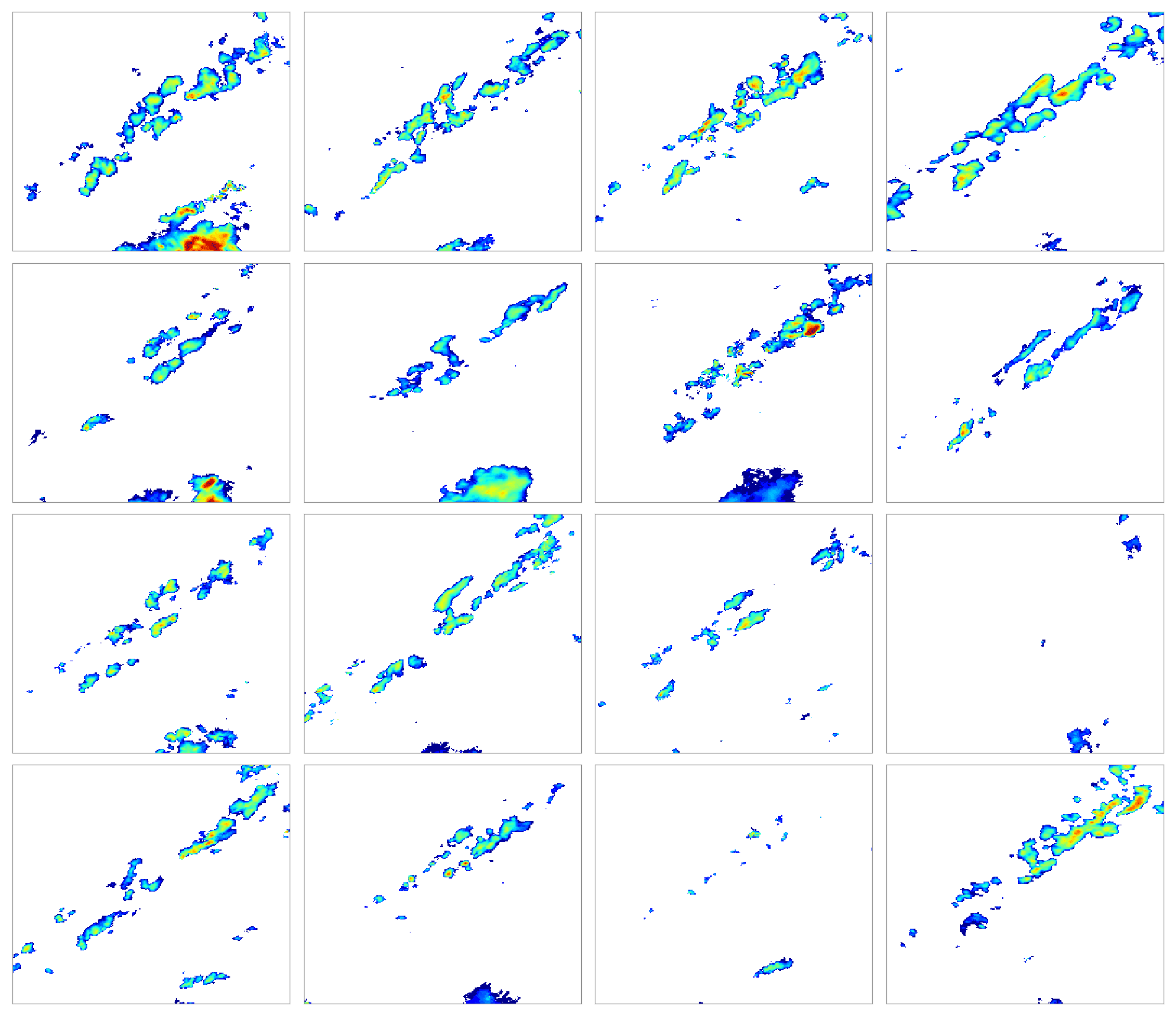}
\caption{Diffusion-model ensemble realisations for the convective case study: sixteen ($4\times4$) independent realisations from the $\varepsilon$-prediction variant. The spread across realisations illustrates the per-sample variability that the ensemble mean averages out.}
\label{fig:ensemble_conv_eps}
\end{figure}

\begin{figure}[!h]
\centering
\includegraphics[width=0.6\textwidth]{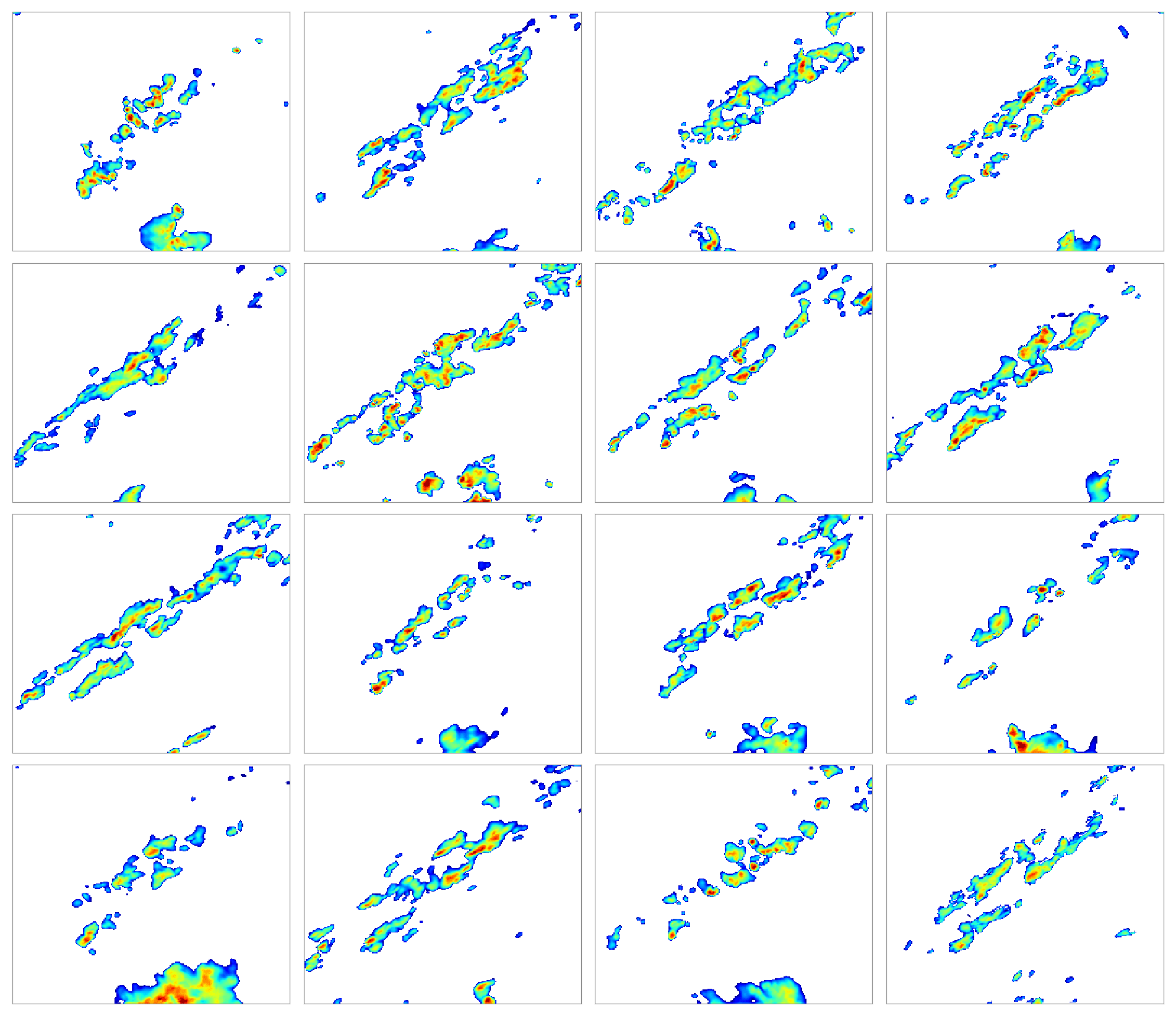}
\caption{Diffusion-model ensemble realisations for the convective case study: sixteen ($4\times4$) independent realisations from the image-prediction variant.}
\label{fig:ensemble_conv_x0}
\end{figure}

\begin{figure}[!h]
\centering
\includegraphics[width=0.4\textwidth]{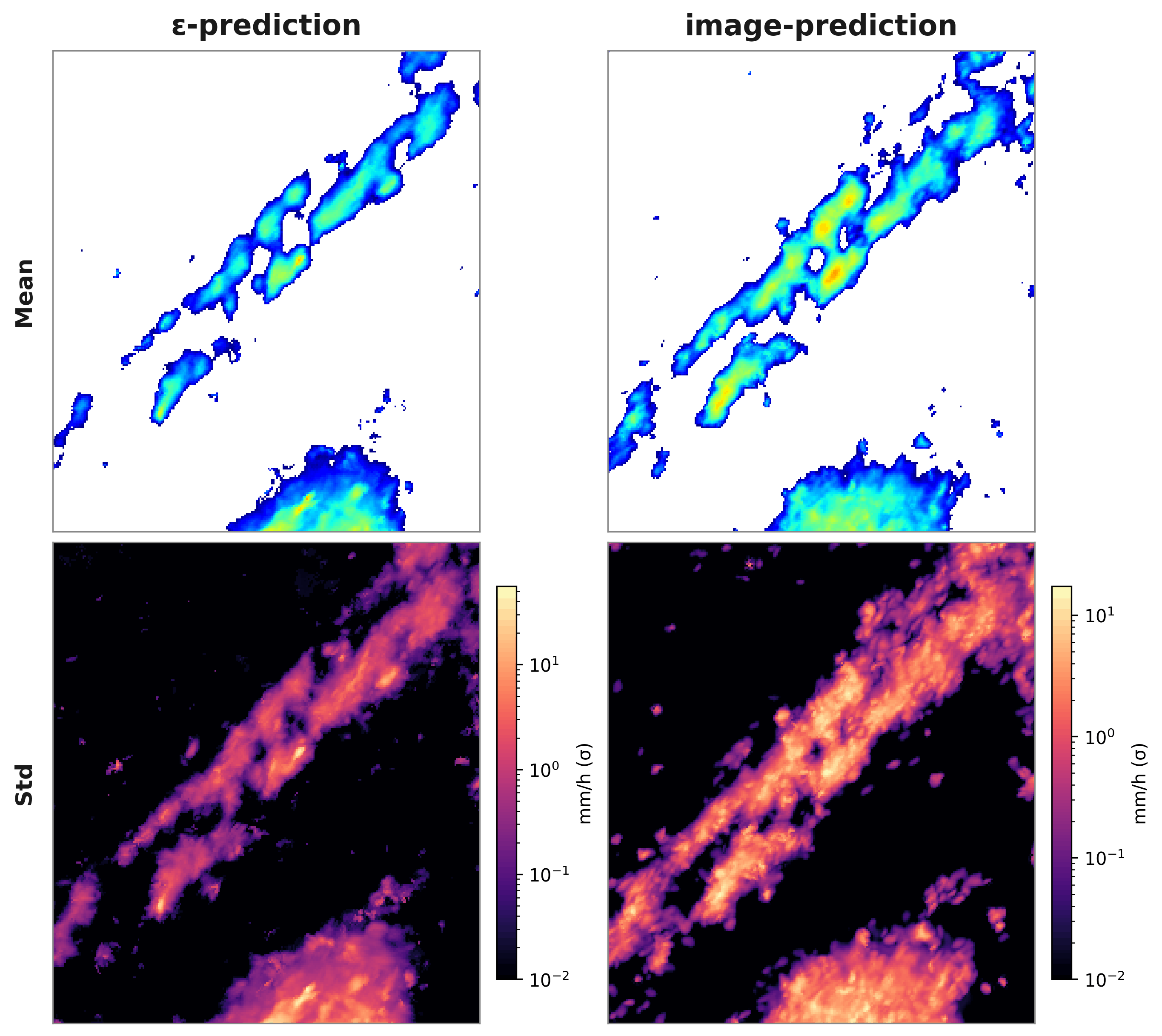}
\caption{Ensemble mean and standard-deviation fields for the convective case study, computed from 50 samples for each diffusion variant.}
\label{fig:ensemble_conv_summary}
\end{figure}

\clearpage

\subsection{Stratiform Event}

\begin{figure}[!h]
\centering
\includegraphics[width=0.6\textwidth]{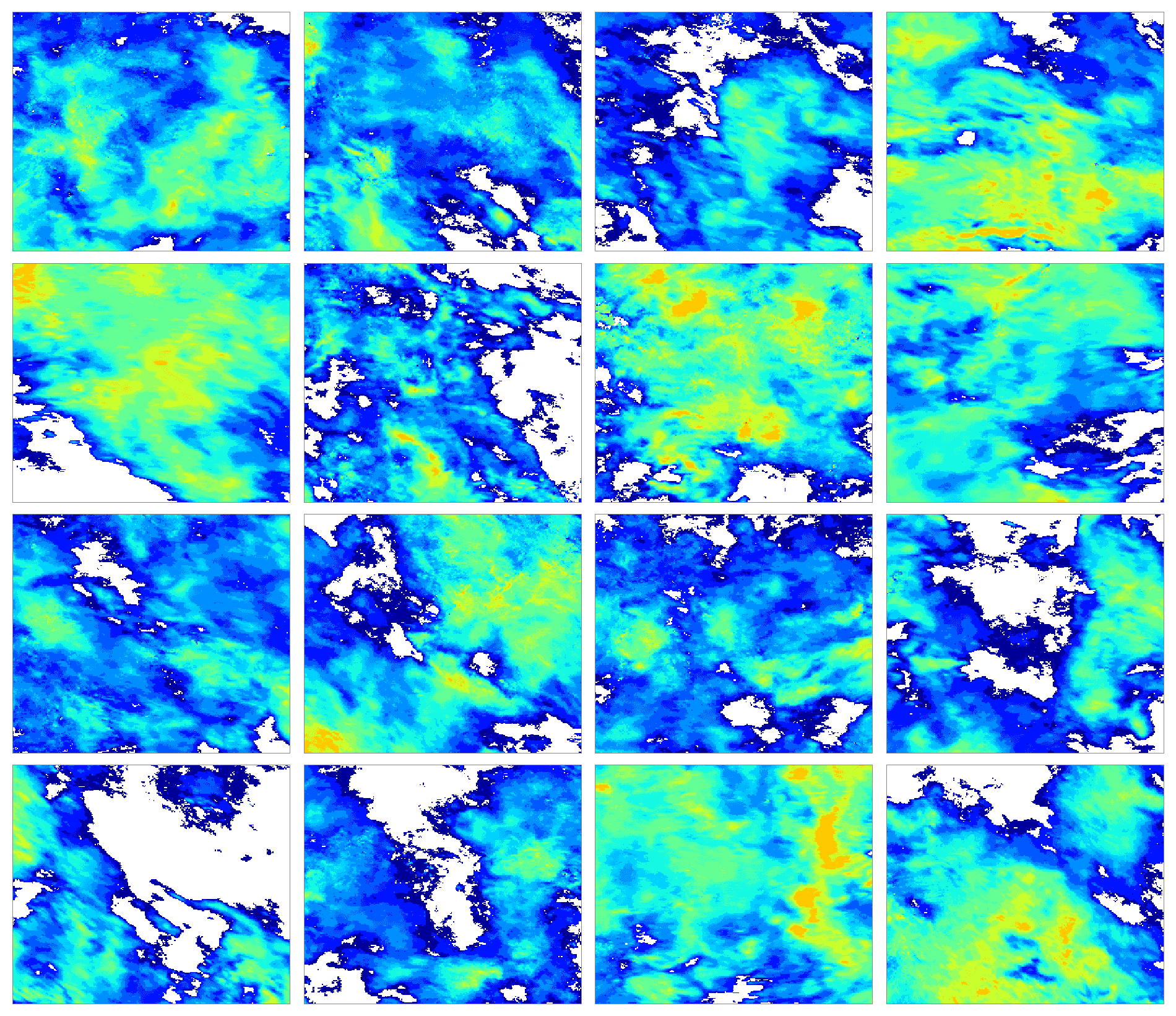}
\caption{Diffusion-model ensemble realisations for the stratiform case study: sixteen ($4\times4$) independent realisations from the $\varepsilon$-prediction variant.}
\label{fig:ensemble_strat_eps}
\end{figure}

\begin{figure}[!h]
\centering
\includegraphics[width=0.6\textwidth]{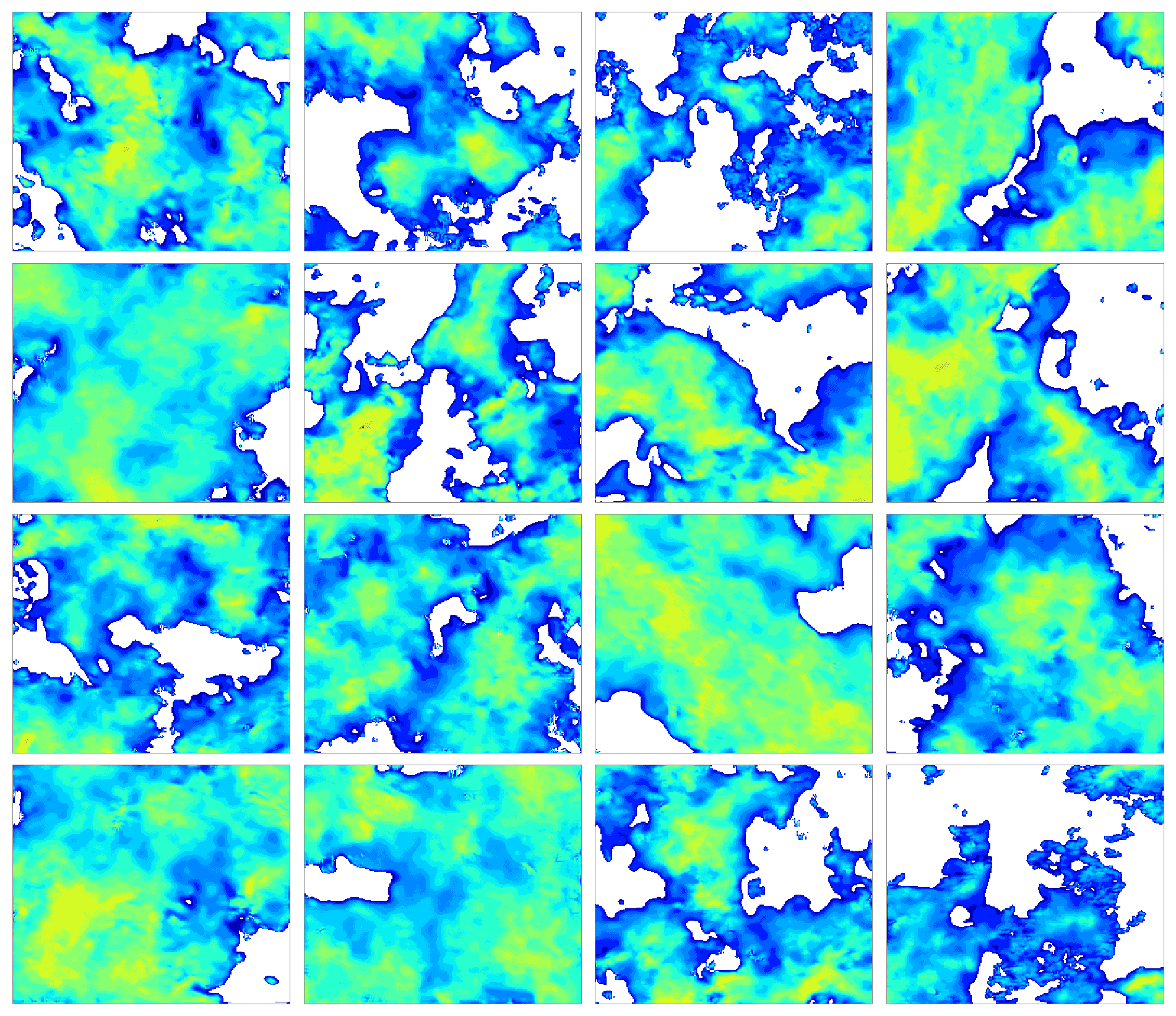}
\caption{Diffusion-model ensemble realisations for the stratiform case study: sixteen ($4\times4$) independent realisations from the image-prediction variant.}
\label{fig:ensemble_strat_x0}
\end{figure}

\begin{figure}[!h]
\centering
\includegraphics[width=0.4\textwidth]{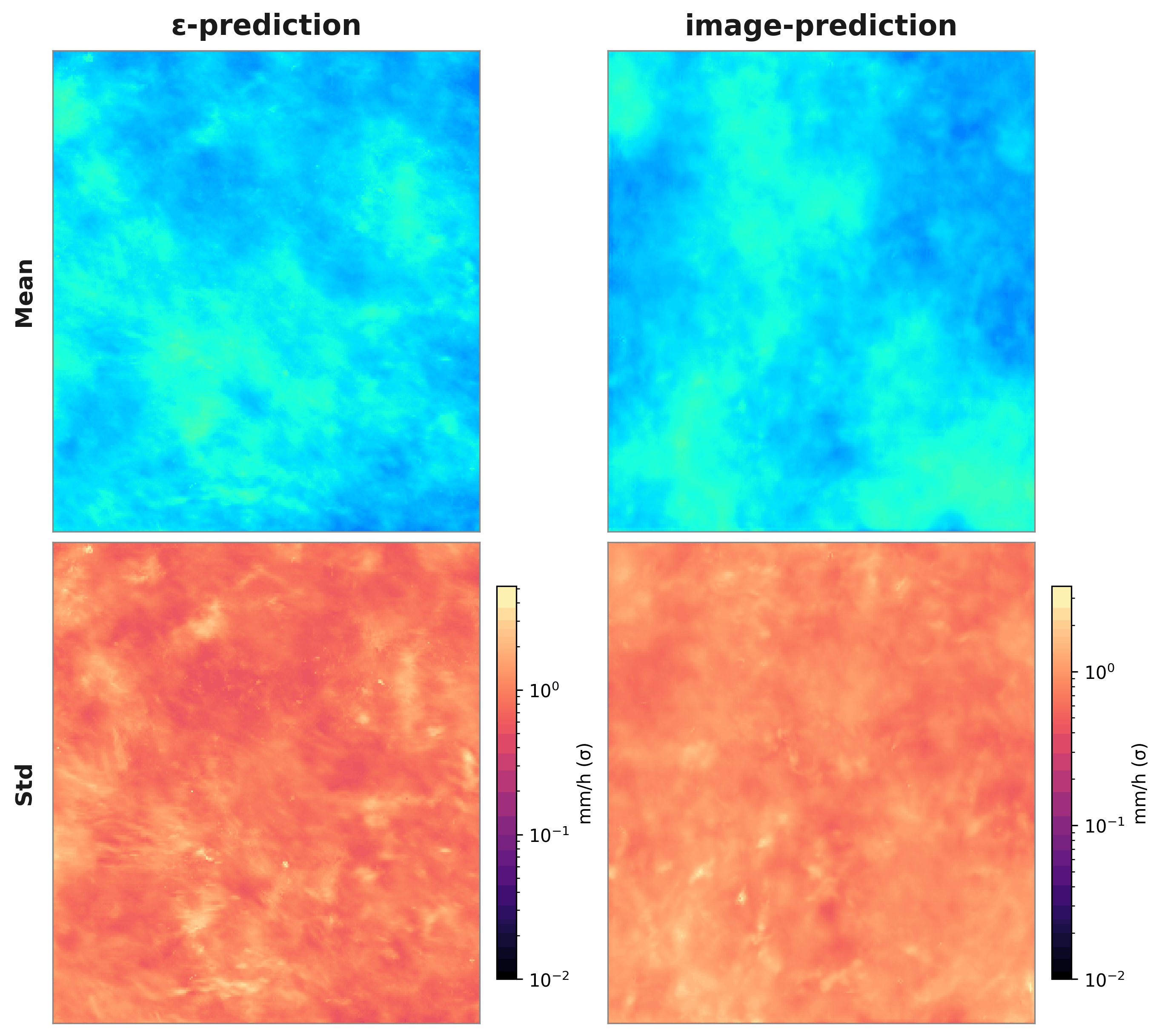}
\caption{Ensemble mean and standard-deviation fields for the stratiform case study, computed from 50 samples for each diffusion variant.}
\label{fig:ensemble_strat_summary}
\end{figure}

\clearpage

\appendix[D]
\appendixtitle{Diffusion ensemble metrics}
\label{appendixD}

The ensemble mean reduces random localisation errors and sampling noise, leading
to more robust detection statistics and an improved representation of low-to-moderate intensity precipitation. However, averaging multiple realisations spatially smooths the precipitation fields and enlarges rainy areas, increasing the occurrence of false alarms. The ensemble mean therefore acts primarily as a robust estimator of the expected precipitation signal, rather than as a mechanism for improving the representation of significant rainfall events (Tables~D1 and~D2).

\input{ensemble_metrics}
\clearpage

\appendix[E]
\appendixtitle{Metrics on the restricted dataset}
\label{appendixE}

\input{binary_restricted}

\input{rmse_restricted}

\clearpage

%



\bibliographystyle{ametsocV6}
\bibliography{references}

\end{document}

%% file: msg_file_format.tex
\begin{table}[htbp]
\centering
\small
\renewcommand{\arraystretch}{1.2}
\newcommand{\cellstack}[1]{\begin{tabular}[t]{@{}l@{}}#1\end{tabular}}
\caption{%
    Overview of SEVIRI Level 1.5 data file format, including the spectral channels
    used in this study, their central wavelengths, and their primary measurement
    purposes. Adapted from EUMETSAT (2017).}
\label{tab:seviri_channels}
\vspace{0.15cm}
\resizebox{0.8\textwidth}{!}{%
\begin{tabular}{|l|l|l|l|l|}
\hline
\cellstack{\textbf{Channel ID}}
& \cellstack{\textbf{Absorption Band}\\\textbf{Channel Type}}
& \cellstack{\textbf{Nominal}\\\textbf{Centre}\\\textbf{Wavelength}\\\textbf{(\textmu m)}}
& \cellstack{\textbf{Spectral}\\\textbf{Bandwidth}\\\textbf{(\textmu m)}}
& \cellstack{\textbf{Spectral}\\\textbf{Bandwidth}\\As \% of energy\\actually detected\\within spectral\\band} \\
\hline
IR 6.2  & \cellstack{Water Vapour\\Core Imager}            & 6.25  & 5.35 to 7.15   & 99.0\,\% \\
\hline
IR 7.3  & \cellstack{Water Vapour\\Pseudo-Sounding}        & 7.35  & 6.85 to 7.85   & 98.0\,\% \\
\hline
IR 8.7  & \cellstack{IR / Window\\Core Imager}             & 8.70  & 8.30 to 9.10   & 98.0\,\% \\
\hline
IR 9.7  & \cellstack{IR / Ozone\\Pseudo-Sounding}          & 9.66  & 9.38 to 9.94   & 99.0\,\% \\
\hline
IR 10.8 & \cellstack{IR / Window\\Core Imager}             & 10.80 & 9.80 to 11.80  & 98.0\,\% \\
\hline
IR 12.0 & \cellstack{IR / Window\\Core Imager}             & 12.00 & 11.00 to 13.00 & 98.0\,\% \\
\hline
IR 13.4 & \cellstack{IR / Carbon Dioxide\\Pseudo-Sounding} & 13.40 & 12.40 to 14.40 & 96.0\,\% \\
\hline
\end{tabular}
}
\end{table}

%% file: classification.tex
\begin{table}[htbp]
\centering
\caption{Pixel-wise rain/no-rain detection metrics for all models at a 0.2 mm/h threshold. Metrics include accuracy, precision, recall, F1 score, critical success index (CSI), and false alarm ratio (FAR). For each metric $\uparrow$ means higher is better, $\downarrow$ means lower is better and best values are shown in bold, second-best values are in italic.}
\begin{tabular}{lcccccc}
\hline
Model & Accuracy ($\uparrow$) & Precision ($\uparrow$) & Recall ($\uparrow$) & F1 ($\uparrow$) & CSI ($\uparrow$) & FAR ($\downarrow$) \\
\hline
Baseline UNet           & 0.82         & 0.64           & 0.76          & \emph{0.69} & \emph{0.53} & 0.36 \\
QuantileUNet:median & \textbf{0.84} & \textbf{0.75} & 0.62             & 0.67       & 0.51      & \textbf{0.25} \\
Multitask UNet       & \emph{0.83} & 0.66            & \emph{0.77} & \textbf{0.71} & \textbf{0.55} & 0.34 \\
DeepLabV3+          & 0.82         & 0.63         & 0.76          & 0.69         & 0.53         & 0.37 \\
SegFormer           & 0.82        & 0.62           & \textbf{0.78} & 0.69       & 0.53       & 0.38 \\
cGAN                & 0.81          & \emph{0.67}   & 0.52          & 0.59        & 0.42       & \emph{0.33} \\
Diffusion (epsilon-prediction) & 0.74 & 0.51 & 0.54 & 0.52 & 0.36 & 0.49 \\
Diffusion (image-prediction) & 0.76  & 0.56 & 0.45 & 0.50 & 0.33 & 0.44 \\
\hline
\end{tabular}
\label{tab:classification_metrics}
\end{table}

%% file: rmse.tex
\begin{table}[h]
\centering
\caption{Pixel-wise root mean square error (RMSE, in mm/h) for all models across different precipitation categories on the test set. Best and second-best values in each column are highlighted in bold and in italic, respectively.}
\resizebox{\textwidth}{!}{%
\begin{tabular}{lccccccc}
\hline
\shortstack{RMSE \\ (mm/h) } & \shortstack{No rain\\$<0.2$} & 
\shortstack{Slight rain\\$[0.2,2.5)$} & 
\shortstack{Moderate rain\\$[2.5,10)$} & 
\shortstack{Heavy rain\\$[10,50)$} & 
\shortstack{Violent rain\\$\ge 50$} & 
\shortstack{Rain\\$\ge 0.2$} & 
\shortstack{All\\$<\infty$} \\
\hline
\textbf{Percentage (\%)} & 73.7 & 21.5 & 4.3 & 0.36 & 0.014 & 26.2 & 100 \\
\hline
Baseline UNet        & 0.36 & \emph{0.92} & 3.35 & 15.36 & 67.96 & 2.88 & \emph{1.50} \\
QuantileUNet:median & \textbf{0.18} & \textbf{0.78} & 3.79 & 16.89 & 75.18 & 3.13 & 1.61 \\
Multitask UNet     & 0.36 & 0.92 & \textbf{3.29} & \emph{15.22} & 67.99 & \textbf{2.85} & \textbf{1.49} \\
DeepLabV3+        & 0.39 & 0.97 & \emph{3.32} & 15.30 & \emph{67.77}& \emph{2.87} & 1.51 \\
SegFormer         & 0.44 & 1.07 & 3.34 & \textbf{14.98} & \textbf{67.25} & 2.88 & 1.52 \\
cGAN              & \emph{0.34} & 1.02 & 4.01 & 16.76 & 71.74 & 3.18 & 1.65 \\
Diffusion (epsilon-prediction)    & 0.71 & 1.40& 4.41 & 17.73  & 78.34 & 3.52 & 1.90 \\
Diffusion (image-prediction)         & 0.63 & 1.37 & 4.21 & 17.24 & 76.06 & 3.41 & 1.83 \\
\hline
\end{tabular}%
}
\label{tab:rmse_metrics}
\end{table}

%% file: bias.tex
\begin{table}[htbp] 
\centering
\caption{Pixel-wise bias (in mm/h) for all models across different precipitation categories on the test set. Best and second-best values in each column are highlighted in bold and italic, respectively.} \resizebox{\textwidth}{!}{%
\begin{tabular}{lccccccc}
\hline
\shortstack{Bias \\ (mm/h) } & 
\shortstack{No rain\\$<0.2$} & 
\shortstack{Slight rain\\$[0.2,2.5)$} &
\shortstack{Moderate rain\\$[2.5,10)$} &
\shortstack{Heavy rain\\$[10,50)$} &
\shortstack{Violent rain\\$\ge 50$} &
\shortstack{Rain\\$\ge 0.2$} & 
\shortstack{All\\$<\infty$} \\ 
\hline 
\textbf{Percentage (\%)} & 73.7 & 21.5 & 4.3 & 0.36 & 0.014 & 26.2 & 100 \\ 
\hline 
Baseline UNet                  & 0.11          & -0.08        & -2.57        & -13.20           & -62.35 & -0.70         & -0.10 \\ 
QuantileUNet:median            & \textbf{0.04} & -0.42         & -3.34       & -15.06           & -70.00             & -1.13      & -0.27 \\ 
Multitask UNet                  & 0.11          & \emph{-0.03} & -2.51       & -13.08            & -62.43            & \emph{-0.65} & \emph{-0.09} \\ 
DeepLabV3+                     &  0.12         & -0.05        & \emph{-2.49 }   & \emph{-13.07} & \emph{-62.31}   & -0.66            & -0.09 \\ 
SegFormer                      & 0.13         & \textbf{0.02} & \textbf{-2.33} & \textbf{-12.54} & \textbf{-61.18}       & \textbf{-0.57} & \textbf{-0.05} \\ 
cGAN                           & \emph{0.06}   &-0.41 & -3.30 & -14.39 & -64.99 & -1.11 & -0.24 \\
Diffusion (epsilon-prediction) & 0.16          & -0.29 & -3.54 & -15.74 & -73.35 & -1.08 & -0.16 \\
Diffusion (image-prediction)   & 0.14          & -0.30 & -3.35 & -15.07 & -70.87 & -1.04 & -0.17 \\ 
\hline \end{tabular}%
} \label{tab:bias}
\end{table}

%% file: mae.tex
\begin{table}[htbp]
\centering
\caption{Pixel-wise Mean Absolute Error (MAE in mm/h) for all models across different precipitation categories on the test set. Best and second-best values in each column are highlighted in bold and in italic, respectively.}
\resizebox{\textwidth}{!}{%
\begin{tabular}{lccccccc}
\hline
\shortstack{MAE \\ (mm/h) } & \shortstack{No rain\\$<0.2$} & 
\shortstack{Slight rain\\$[0.2,2.5)$} & 
\shortstack{Moderate rain\\$[2.5,10)$} & 
\shortstack{Heavy rain\\$[10,50)$} & 
\shortstack{Violent rain\\$\ge 50$} & 
\shortstack{Rain\\$\ge 0.2$} & 
\shortstack{All\\$<\infty$} \\
\hline
\textbf{Percentage (\%)} & 73.7 & 21.5 & 4.3 & 0.36 & 0.014 & 26.2 & 100 \\
\hline
Baseline UNet                        & 0.11 & \emph{0.62} & 2.80 & 13.43 & 62.45 & \emph{1.19} & 0.40 \\
QuantileUNet:median                  & \textbf{0.04} & \textbf{0.59} & 3.35 & 15.08 & 70.00 & 1.28 & \textbf{0.37} \\ 
Multitask UNet                        & 0.11 & 0.63 & \emph{2.73} & \emph{13.29} & 62.50 & \textbf{1.18} & \emph{0.39} \\
DeepLabV3+                           & 0.12 & 0.64 & 2.75 & 13.33 & \emph{62.37} & 1.20 & 0.40 \\
SegFormer                            & 0.14 & 0.68 & \textbf{2.70} & \textbf{12.98} & \textbf{61.65} & 1.21 & 0.42 \\
cGAN                                 & \emph{0.07} & 0.69 & 3.46 & 14.84 & 65.70 & 1.37 & 0.41 \\
Diffusion (epsilon-pred)             & 0.16 & 0.79 & 3.74 & 15.84 & 73.36 & 1.52 & 0.52 \\
Diffusion (image-prediction)         & 0.14 & 0.83 & 3.62 & 15.33 & 70.91 & 1.52 & 0.50 \\
\hline
\end{tabular}%
}
\label{tab:MAE}
\end{table}

%% file: pearson.tex
\begin{table}[htbp]
\centering
\caption{Pixel-wise Pearson correlation for all models across different precipitation categories on the test set. Best and second-best values in each column are highlighted in bold and in italic, respectively.}
\resizebox{\textwidth}{!}{%
\begin{tabular}{lccccccc}
\hline
\shortstack{Pearson} & \shortstack{No rain\\$<0.2$} & 
\shortstack{Slight rain\\$[0.2,2.5)$} & 
\shortstack{Moderate rain\\$[2.5,10)$} & 
\shortstack{Heavy rain\\$[10,50)$} & 
\shortstack{Violent rain\\$\ge 50$} & 
\shortstack{Rain\\$\ge 0.2$} & 
\shortstack{All\\$<\infty$} \\
\hline
\textbf{Percentage (\%)} & 73.7 & 21.5 & 4.3 & 0.36 & 0.014 & 26.2 & 100 \\
\hline
Baseline Unet                        & 0.18 & 0.28 & 0.17 & 0.25 & \emph{0.26} & \emph{0.43} & \emph{0.51} \\
QuantileUNet:median                  & \emph{0.18} & \textbf{0.29} & 0.15 & 0.18 & 0.08 & 0.36 & 0.47 \\ 
Multitask UNet                        & \textbf{0.19} & \emph{0.29} & \textbf{0.18} & \textbf{0.27} & 0.25 & \textbf{0.44} & \textbf{0.52} \\
DeepLabV3+                           & 0.17 & 0.28 & \emph{0.18} & 0.24 & \textbf{0.30}& 0.43 & \emph{0.51} \\
SegFormer                            & 0.17 & 0.27 & \emph{0.17} & \emph{0.25} & 0.20 & 0.42 & 0.49 \\
cGAN                                 & 0.10 & 0.18 & 0.10 & 0.18 & 0.11 & 0.30 & 0.36 \\
Diffusion (epsilon-pred)             & 0.07 & 0.06 & 0.02 & 0.01 & -0.05 & 0.05 & 0.11 \\
Diffusion (image-prediction)         & 0.08 & 0.09 & 0.06 & 0.09 & 0.08 & 0.13 & 0.19 \\
\hline
\end{tabular}%
}
\label{tab:pearson}
\end{table}

%% file: implementation_unets.tex
\begin{table}[htbp]
\caption{Architecture and training configuration for the U-Net family. All three models share an 
identical encoder--decoder backbone and are trained with 128$\times$128 crops, random horizontal 
and vertical flips, and the Adam optimizer with a cosine-annealing-with-warm-restarts 
schedule ($T_0=100$, $T_\mathrm{mult}=2$, $\eta_\mathrm{min}=10^{-6}$) starting from an initial 
learning rate of $10^{-3}$. For the Multitask U-Net, $\omega=10$ balances the regression and 
segmentation losses. Rows in the lower block list the settings that differ across variants.}
\label{tab:unet_details}
\centering
\begin{tabular}{lccc}
\hline
 & U-Net & Multitask U-Net & Quantile U-Net \\
\hline
\multicolumn{4}{l}{\textit{Shared architecture}} \\
\quad Encoder channels    & \multicolumn{3}{c}{[32, 64, 128, 256]} \\
\quad Bottleneck channels & \multicolumn{3}{c}{512} \\
\quad Decoder upsampling   & \multicolumn{3}{c}{Transpose conv.} \\
\quad Dropout             & \multicolumn{3}{c}{0.05} \\
\hline
\multicolumn{4}{l}{\textit{Variant-specific settings}} \\
\quad Output heads          & 1 (regression) & 2 (reg.\ + seg.) & 9 (deciles) \\
\quad Loss                  & MSE            & MSE $+\,\omega\,$BCE & Quantile loss \\
\quad Target transformation & $\varepsilon$-zero-noise$^{a}$ & $\varepsilon$-zero-noise$^{a}$ & Log-zero-noise$^{b}$ \\
\quad Batch size            & 16             & 16               & 32 \\
\quad Best epoch            & 290            & 385              & 260 \\
\hline
\end{tabular}

\smallskip
{\footnotesize
$^{a}$\,A small uniform noise $\varepsilon \sim \mathcal{U}(-\epsilon, +\epsilon)$
is added to the precipitation target exclusively at zero-rain pixels,
preventing the model from collapsing to a trivial all-zero prediction.\\
$^{b}$\,The precipitation target undergoes the same transformation as in \cite{Amell2025}.}
\end{table}

%% file: implementation_segformer_deeplav3+.tex
\begin{table}[htbp]
\caption{Architecture and training configuration for the DeepLabV3+ and SegFormer models. Both are 
trained with 224$\times$224 crops; training with 128$\times$128 crops degraded performance. 
Both use random horizontal and vertical flips, the Adam optimizer 
($\eta_0=10^{-3}$), and a cosine-annealing-with-warm-restarts schedule ($T_0=100$, 
$T_\mathrm{mult}=1$, $\eta_\mathrm{min}=10^{-6}$).}
\label{tab:deeplab_segformer_details}
\centering
\begin{tabular}{lcc}
\hline
 & DeepLabV3+ & SegFormer \\
\hline
\multicolumn{3}{l}{\textit{Encoder / backbone}} \\
\quad Type & ResNet-50 (no pretraining) & Hierarchical transformer (MiT-B2) \\
\hline
\multicolumn{3}{l}{\textit{Decoder / head}} \\
\quad Context module & Lightweight ASPP         & MLP decoder \\
\quad Dilation rates & $(1,2,4,6)$              & --- \\
\quad Output head    & $1\times1$ conv, $n{=}1$ & $1\times1$ conv, $n{=}1$ \\
\hline
\multicolumn{3}{l}{\textit{Training}} \\
\quad Loss                       & MSE                 & MSE \\
\quad Target transformation & $\varepsilon$-zero-noise$^{a}$ & $\varepsilon$-zero-noise$^{a}$ \\
\quad Batch size                 & 16                  & 16 \\
\quad Best epoch                 & 283                 & 286  \\
\hline
\end{tabular}

\smallskip
{\footnotesize $^{a}$\,A small uniform noise $\varepsilon \sim \mathcal{U}(-\epsilon, +\epsilon)$
is added to the precipitation target exclusively at zero-rain pixels,
preventing the model from collapsing to a trivial all-zero prediction.}
\end{table}

%% file: implementation_generative.tex
\begin{table}[htbp]
\centering
\small
\setlength{\tabcolsep}{9pt}
\renewcommand{\arraystretch}{1.3}
\caption{%
    Implementation details for the generative models.
    All models were trained using random crops of size 128$\times$128.
    All channels undergo z-score normalization, $(x-\mu)/(k\sigma)$, where $\mu$ and $\sigma$ are
    the training-set mean and standard deviation. Inputs use $k=3$ for all models.
    BL: bilinear interpolation; H/V-flip: random horizontal and vertical flips;
    EMA: exponential moving average of model weights.
}
\label{tab:implementation_generative}
\resizebox{\textwidth}{!}{%
\begin{tabular}{@{} l ccc @{}}
\hline
& \textbf{cGAN}$^\star$
& \textbf{DDPM} ($\varepsilon$-pred.)
& \textbf{DDPM} (image-pred.) \\
\hline
\multicolumn{4}{@{}l}{\textit{Architecture}} \\
\quad Generator / denoiser
    & U-Net $[32,64,128,256]$
    & Attn.\ U-Net ($c{=}64$, mult.\ $[1,1,2,3,4]$)
    & Attn.\ U-Net ($c{=}64$, mult.\ $[1,1,2,3,4]$) \\
\quad Bottleneck channels   & 512 & 256 & 256 \\
\quad Residual blocks / level & --- & 2 & 2 \\
\quad Attention resolution  & --- & $16{\times}16$ & $16{\times}16$ \\
\quad Discriminator         & PatchGAN (3 conv.\ layers, 64 base filters) & --- & --- \\
\quad Upsampling            & BL & Learned & Learned \\
\quad Dropout               & 0.05 & 0.10 & 0.10 \\
\quad Prediction target     & --- & $\varepsilon$ & image \\
\quad EMA                   & --- & --- & 0.9999 \\
\hline
\multicolumn{4}{@{}l}{\textit{Training objective}} \\
\quad Generator loss
    & $\lambda_\mathrm{L1}\,\mathrm{L1} + \mathrm{LSGAN}$ ($\lambda_\mathrm{L1}{=}10$)
    & MSE & MSE \\
\quad Discriminator loss    & LSGAN & --- & --- \\
\quad Diffusion schedule    & --- & Cosine ($T{=}500$) & Cosine ($T{=}500$) \\
\hline
\multicolumn{4}{@{}l}{\textit{Optimization}} \\
\quad Optimizer
    & Adam ($\beta_1{=}0.5$, $\beta_2{=}0.999$) & AdamW & AdamW \\
\quad Learning rate         & $2{\times}10^{-4}$ & $10^{-4}$ & $10^{-4}$ \\
\hline
\multicolumn{4}{@{}l}{\textit{Data pipeline}} \\
\quad Input channels        & 4 & 7 & 7 \\
\quad Target normalization  & --- & z-score ($k{=}1$) & z-score ($k{=}1$) \\
\quad Batch size            & 32 & 16 & 16 \\
\hline
\multicolumn{4}{@{}l}{\textit{Training runtime}} \\
\quad Best epoch / steps    & 283 & 723{,}000 & 822{,}000 \\
\hline
\end{tabular}%
}

\vspace{6pt}
{\footnotesize    $^\star$The cGAN uses 4 input channels only: 6.2\,\textmu m, 8.7\,\textmu m, 
    10.8\,\textmu m, and 12.0\,\textmu m.
    Training with the full 7-channel set yielded no improvement in pixel-wise metrics; 
    we speculate that the richer input may give the discriminator an advantage that 
    destabilises the adversarial training dynamic.}
\end{table}

%% file: ensemble_metrics.tex
\begin{table}[htbp]
\centering
\caption{Pixel-wise binary rain/no-rain detection metrics computed from the ensemble mean of 50 generated realisations for both diffusion model variants, using a rainfall threshold of 0.2 mm/h. Metrics include accuracy, precision, recall, F1 score, critical success index (CSI), and false alarm ratio (FAR).}
\begin{tabular}{lcccccc}
\hline
Model & Accuracy ($\uparrow$) & Precision ($\uparrow$) & Recall ($\uparrow$) 
      & F1 ($\uparrow$) & CSI ($\uparrow$) & FAR ($\downarrow$) \\
\hline
Diffusion ($\epsilon$-prediction)         & 0.75 & 0.51 & 0.83 &0.64 &0.47 &0.49 \\
Diffusion (image-prediction)         & 0.78     & 0.56 &   0.75 &0.64 &0.47 & 0.44 \\
\hline
\end{tabular}
\label{tab:binary_ensembles}
\end{table}

\begin{table}[htbp]
\centering
\caption{RMSE (mm/h) computed from the ensemble mean of 50 generated realisations for the two diffusion model variants. Errors are reported separately for each rainfall-intensity category, as well as for all rainy pixels ($\geq 0.2$ mm/h) and the complete dataset. The percentage row indicates the proportion of pixels belonging to each category in the test set.}
\begin{tabular}{lccccccc}
\hline
\shortstack{RMSE \\ (mm/h)} &
  \shortstack{No rain\\$<0.2$} &
  \shortstack{Slight rain\\$[0.2,2.5)$} &
  \shortstack{Moderate rain\\$[2.5,10)$} &
  \shortstack{Heavy rain\\$[10,50)$} &
  \shortstack{Violent rain\\$\ge 50$} &
  \shortstack{Rain\\$\ge 0.2$} &
  \shortstack{All\\$<\infty$} \\
\hline
\textbf{Percentage (\%)} & 73.80 &  21.54 &4.29 & 0.36 & 0.01 & 26.2 & 100 \\
\hline
Diffusion ($\epsilon$-prediction)               & 0.31 & 0.73 & 3.96 &17.55 &78.10 &3.24 &1.68\\
Diffusion (image-prediction)    & 0.30 & 0.78 & 3.85 & 16.93 & 75.61 &3.15 &1.64 \\
\hline
\end{tabular}%
\label{tab:rmse_ensembles}
\end{table}

%% file: binary_restricted.tex
\begin{table}[h]
\centering
\caption{Binary detection metrics (threshold = 0.2~mm\,h$^{-1}$) for the deterministic models retrained on the restricted dataset used for the generative models. The test set is unchanged.}
\begin{tabular}{lcccccc}
\hline
Model & Accuracy ($\uparrow$) & Precision ($\uparrow$) & Recall ($\uparrow$) 
      & F1 ($\uparrow$) & CSI ($\uparrow$) & FAR ($\downarrow$) \\
\hline
Baseline UNet     & 0.82 & 0.64 & 0.74 & \emph{0.69} & \emph{0.52} & 0.36 \\
QuantileUNet:median & \textbf{0.84} & \textbf{0.75} & 0.61 & 0.67 & 0.51 & \textbf{0.25} \\
Multitask UNet    & \emph{0.83} & \emph{0.66} & \textbf{0.76} & \textbf{0.71} & \textbf{0.55} & \emph{0.34} \\
DeepLabV3+        & 0.79 & 0.58 & 0.76& 0.66 & 0.49 & 0.42 \\
SegFormer         & 0.82 & 0.62 & \emph{0.76} & 0.69 & 0.52 & 0.38 \\
\hline
\end{tabular}
\label{tab:binary_restricted}
\end{table}

%% file: rmse_restricted.tex
\begin{table}[h]
\centering
\caption{RMSE (mm\,h$^{-1}$) for the deterministic models retrained on the restricted dataset used for the generative models. The test set is unchanged.}
\resizebox{\textwidth}{!}{%
\begin{tabular}{lccccccc}
\hline
\shortstack{RMSE \\ (mm/h)} &
  \shortstack{No rain\\$<0.2$} &
  \shortstack{Slight rain\\$[0.2,2.5)$} &
  \shortstack{Moderate rain\\$[2.5,10)$} &
  \shortstack{Heavy rain\\$[10,50)$} &
  \shortstack{Violent rain\\$\ge 50$} &
  \shortstack{Rain\\$\ge 0.2$} &
  \shortstack{All\\$<\infty$} \\
\hline
\textbf{Percentage (\%)} & 73.80 &  21.54 &4.29 & 0.36 & 0.01 & 26.2 & 100 \\
\hline
Baseline UNet  & \emph{0.35} & 0.93 & 3.44 & 15.62 & 70.23   & 2.94 & 1.54\\
QuantileUNet:median  & \textbf{0.18} & \textbf{0.79} & 3.78 & 16.88 & 74.86  & 3.13 & 1.61 \\
Multitask UNet & 0.36 & \emph{0.92} & \emph{3.35} & \emph{15.40}   & \emph{69.18} & \emph{2.89} & \textbf{1.51} \\
DeepLabV3+     & 0.44 & 0.96 & 3.55 & 15.93 & 70.29  & 2.99 & 1.58 \\
SegFormer      & 0.42 & 1.01 & \textbf{3.33} & \textbf{15.18} & \textbf{66.78}  & \textbf{2.87} &  \emph{1.51}\\
\hline
\end{tabular}%
}
\label{tab:rmse_restricted}
\end{table}